\documentclass[aps,twocolumn,showpacs,email,amssymb,nobibnotes]{revtex4-1}
\usepackage{graphicx,color}

\usepackage{amssymb}
\usepackage{amsbsy}
\usepackage{amsthm,mathrsfs,amsopn}
\usepackage{amsmath}
\usepackage{verbatim}
\usepackage{graphicx}
\usepackage{subfigure}
\usepackage{bm}
\usepackage{url}

\begin{document}

\title{Suppressing escape events in maps of the unit interval with demographic noise}

\author{C\'{e}sar Parra-Rojas$^{1}$, Joseph D. Challenger$^{2}$, Duccio Fanelli$^{3}$, Alan J. McKane$^{1}$}
\affiliation{
$^1$Theoretical Physics Division, School of Physics and Astronomy, The University of Manchester, Manchester M13 9PL, United Kingdom\\
$^2$Department of Infectious Disease Epidemiology, Imperial College London, London, W2 1PG, UK \\
$^3$Dipartimento di Fisica e Astronomia, Universit\`{a} di Firenze, INFN and CSDC, Via Sansone 1, 50019 Sesto Fiorentino, Firenze, Italy
}

\begin{abstract} 
We explore the properties of discrete-time stochastic processes with a bounded state space, whose deterministic limit is given by a map of the unit interval. We find that, in the mesoscopic description of the system, the large jumps between successive iterates of the process can result in probability leaking out of the unit interval, despite the fact that the noise is multiplicative and vanishes at the boundaries. By including higher-order terms in the mesoscopic expansion, we are able to capture the non-Gaussian nature of the noise distribution near the boundaries, but this does not preclude the possibility of a trajectory leaving the interval. We propose a number of prescriptions for treating these escape events, and we compare the results with those obtained for the metastable behavior of the microscopic model, where escape events are not possible. We find that, rather than truncating the noise distribution, censoring this distribution to prevent escape events leads to results which are more consistent with the microscopic model. The addition of higher moments to the noise distribution does not increase the accuracy of the final results, and it can be replaced by the simpler Gaussian noise.
\end{abstract}


\maketitle

\section{Introduction}
\label{sec:intro}
Simple discrete-time models are frequently used to gain insight into the behavior of more complex models of real world systems. The logistic map, for example, can be used to study the population dynamics of species that have nonoverlapping generations. Here, the many interactions between the individuals in the population (birth, death, competition etc.) are replaced with an effective interaction which returns the population of the next generation as a function of the current one. More theoretically, this system also provides a very simple setting in which to study deterministic chaos \cite{ott_93,strogatz_94}.

The effect of noise on such a system is interesting for a number of reasons. Any real world population will be subject to internal (demographic) and external (environmental) fluctuations. From a more theoretical point of view, the inclusion of noise allows us to consider a simple system which contains two types of uncertainty: it exhibits chaotic and stochastic behavior (see e.g. Refs~\cite{crutchfield_82,gao_99}). In a series of articles \cite{challenger_13,challenger_14,parra_14}, we introduced a microscopic model from which the logistic map is recovered in the thermodynamic limit (the limit of zero fluctuations). This process is frequently studied in continuous-time models, but to the best of our knowledge had not been carried out in discrete-time systems. From this microscopic model, expressed as a Markov chain, we established a mesoscopic formalism, to elucidate certain aspects of the process, and also to connect with well-known techniques developed for continuous-time stochastic models. This mesoscopic description resembles the deterministic map, but has an extra noise term. We stress that the form of this noise is derived from the microscopic model, and is not simply `put in by hand', as in previous studies. We also extended the formalism from one to many variables, using a predator-prey model as an example~\cite{parra_14}. 

Nonlinear models, in both discrete and continuous time, have been used to explain phenomena such as periodic or aperiodic oscillations found in time series data. Such time series can be measured in a wide range of physical and biological systems. Ecological systems provide a particularly interesting framework to study nonlinear effects, both in laboratory and natural settings (see e.g. Refs.~\cite{costantino_95,rohani_96,costantino_97,beninca_15}). Although microscopic models were not developed in these studies, these groups appreciated that stochasticity---whether viewed as intrinsic or extrinsic---had to be introduced into the model to facilitate comparison with the data.

Discrete-time models---in the deterministic limit---are commonly defined on a fixed interval. For instance, the dynamics of the one-dimensional logistic map is confined to the interval between 0 and 1. The principal reason for this is to facilitate its study; even with only one variable, discrete-time systems are capable of such a wide range of behavior (e.g., cycles, chaos, intermittency) that they allow for large jumps between successive time points (or, from a population biology point of view, `generations'). This should be contrasted with a system of ordinary differential equations which has trajectories that take the form of continuous flows. As a result of these large jumps, the addition of stochasticity can in principle lead to trajectories leaving the interval on which the dynamics is defined. This may be unphysical in that it leads to either negative population numbers or to the population blowing up. Equally, if one wants to study the effects of noise on the attractor of the underlying deterministic map, one is not so interested in trajectories that leave the interval. 

A variety of approaches have been taken in the past to deal with this problem. In the physics literature, it is common to study the system in the weak noise regime (e.g.,~Refs.~\cite{crutchfield_82,fogedby_05}). This means that fluctuations are highly unlikely to cause a trajectory to leave the interval, and so this possibility may be neglected quite safely in applications. One can also choose to condition the noise to explicitly forbid escape events~\cite{haken_81}. Alternatively, one can treat the escape events as part of the dynamics of the system and study them statistically (see, e.g.,~Refs.~\cite{erguler_08,demaeyer_09}). In the mathematical literature, on the other hand, one can view the desired behavior (i.e., trajectories that remain in the interval for the duration of a simulation) as metastable behavior---see for example Ref.~\cite{faure_14}, and references therein. The interpretation in this case is that, if the system is allowed to evolve for a sufficiently long period of time, all of the probability distribution will `leak' out of the interval, representing the true, stationary behavior of the system. 

The microscopic model used as a starting point for our investigations does not suffer from these problems. This is because the process is defined as a Markov chain on a one-dimensional, finite lattice, and the transition probabilities that specify the jump from one lattice point to another are such that transitions outside the interval are not allowed. In the mesoscopic description, however, the discrete lattice points are replaced with a continuous variable (details will be given in Sec.~\ref{sec:meso}). In this case, it does become possible, when the fluctuations are large, for probability to leak out of the interval. Our previous investigations did not focus on the point but, when studying the process for very strong noise, this effect becomes noticeable. Since the mesoscopic description should be consistent with the microscopic model, this probability leakage must be forbidden. To state it differently, the true microscopic model --- to which the mesoscopic model is an approximation --- has zero probability outside the relevant interval, and the mesoscopic description needs to properly reflect this fact. In this paper, we will examine the issue of probability leakage out of the interval from a number of perspectives and, through a more careful analysis, describe and compare several methods for remedying this problem.

The outline of the paper is as follows. In Sec.~\ref{sec:micro} we will discuss the microscopic starting point: a Markov chain with states $n=0,1,\ldots,N$. To illustrate the problem we use a model that leads to the logistic map in the deterministic limit, although the ideas described here will apply more generally. We examine the nature of the eigenvalues and eigenvectors of the transition matrix of this model for small to moderate values of $N$ (up to a few hundred) in order to compare with the results found in the mesoscopic approximation. In Sec.~\ref{sec:meso} we briefly present the formalism used in our previous work and provide results of the mesoscopic expansion of the Markov chain model. We will also show that, unlike the microscopic model, the mesoscopic system allows probability to leave the system and explore this aspect quantitatively. This allows us to develop the ideas which lead us to a prescription ensuring that trajectories do not leave the interval on which the process is defined. This is discussed in Sec.~\ref{sec:trun_cen}, while results obtained from the methods proposed in this section are presented in Sec.~\ref{sec:results}. We discuss our findings in Sec.~\ref{sec:discuss} and give some brief technical details in the appendices.

\section{Microscopic description}
\label{sec:micro}
Our starting point is a microscopic model that has the form of a Markov chain, which we write as 
\begin{equation}
P_{n, t+1}=\sum^{N}_{m=0}Q_{nm}P_{m, t},
\label{markov_chain}
\end{equation}
where $P_{n, t}$ denotes the probability for the system to be found in state $n$ at time $t$. The system is defined on a one-dimensional lattice with $N+1$ sites and is updated by an application of the transition matrix $\bm{Q}$, once an initial condition is provided. Our arguments will be quite general, but for numerical illustration we will focus on models that have the following binomial form:
\begin{equation}
Q_{nm}= { N \choose n} \left[ p\left( \frac{m}{N}\right) \right]^{n}\,\left[ 1 - p\left( \frac{m}{N} \right) \right]^{N-n},
\label{binomial}
\end{equation}
which was used in our previous work~\cite{challenger_13,challenger_14}. Here we have introduced a function $p$ such that $0\leq p\left( m/N \right) \leq 1$. We can find the equation of the evolution of the first moment $\langle n \rangle = \sum_n n P_n$, as follows:
\begin{equation}
\langle n_{t+1} \rangle = \sum_n n P_{n,t+1} = \sum_n \sum_m n Q_{nm} P_{m,t},
\end{equation}
and since $Q_{nm}$ is a binomial distribution, so that, $\sum_n n Q_{nm}= Np(m/N)$, this leads to 
\begin{equation}
\langle n_{t+1} \rangle = \sum_m Np\left(\frac{m}{N}\right) P_{m,t} = N\left\langle p\left(\frac{n_t}{N}\right) \right\rangle. 
\label{first_moment}
\end{equation}
The macroscopic evolution equation is therefore $z_{t+1}=p(z_t)$, where $z_t=\lim_{N\to \infty} \langle n_t \rangle /N$. Choosing the function $p$ to be $p(z)=\lambda z(1-z)$, we see that the model corresponds to the logistic map in the deterministic limit. We note that in this case $p(0)=0$ and $p(1)=0$, which in terms of the transition matrix implies that $Q_{n 0} = Q_{n N} = 0$ for $n > 0$, with $Q_{00}$ and $Q_{0 N}$ not determined. Since $\bm{Q}$ is a stochastic matrix, that is $\sum_{n}Q_{nm} = 1$, for all $m$, we define $Q_{00}=1$ and $Q_{0 N}=1$. In terms of a population model these correspond, respectively, to an extinct population remaining extinct and a population which has reached a critical size $N$ becoming extinct. We would expect both of these conditions to hold in a wide range of population models.

The solution to the Markov chain Eq.~\eqref{markov_chain} is given by
\begin{equation}
P_{n,t} = \sum\limits_{i,m=0}^N v^{(i)}_n\left(\rho^{(i)}\right)^t w^{(i)}_m P_{m,0},
\label{MC_soln}
\end{equation}
where $\rho^{(i)}$ is the $i$th eigenvalue of the transition matrix $\bm{Q}$ and $\bm w^{(i)}$ and $\bm v^{(i)}$ are the corresponding left and right eigenvectors. Thus the long-time, and even moderate-time, behavior of the system will be dominated by the largest eigenvalues of $\bm{Q}$. We now explore the qualitative and quantitative nature of these eigenvalues and associated eigenvectors for later comparison with the results found from the mesoscopic form of the model.

Since $\bm{Q}$ is a stochastic matrix, it follows that it has an eigenvalue equal to 1, with a corresponding left eigenvector given by $(1\,1\,\ldots\,1)^{\rm T}$. From the general theory of Markov chains~\cite{feller_68,cox_65}, we would expect this to correspond to the stationary state and for this eigenvalue to be the largest. The properties of $Q_{n 0}$, described above as following from the condition $p(0)=0$, imply that the boundary at $n=0$ is absorbing, or in terms of a population model, extinction is the final state of the system with probability $1$. Therefore we would expect the right eigenvector corresponding to the eigenvalue $1$ to be $(1\,0\,\ldots\,0)^{\rm T}$, which is straightforward to verify.

\begin{figure}
\flushleft
\includegraphics[scale=.6]{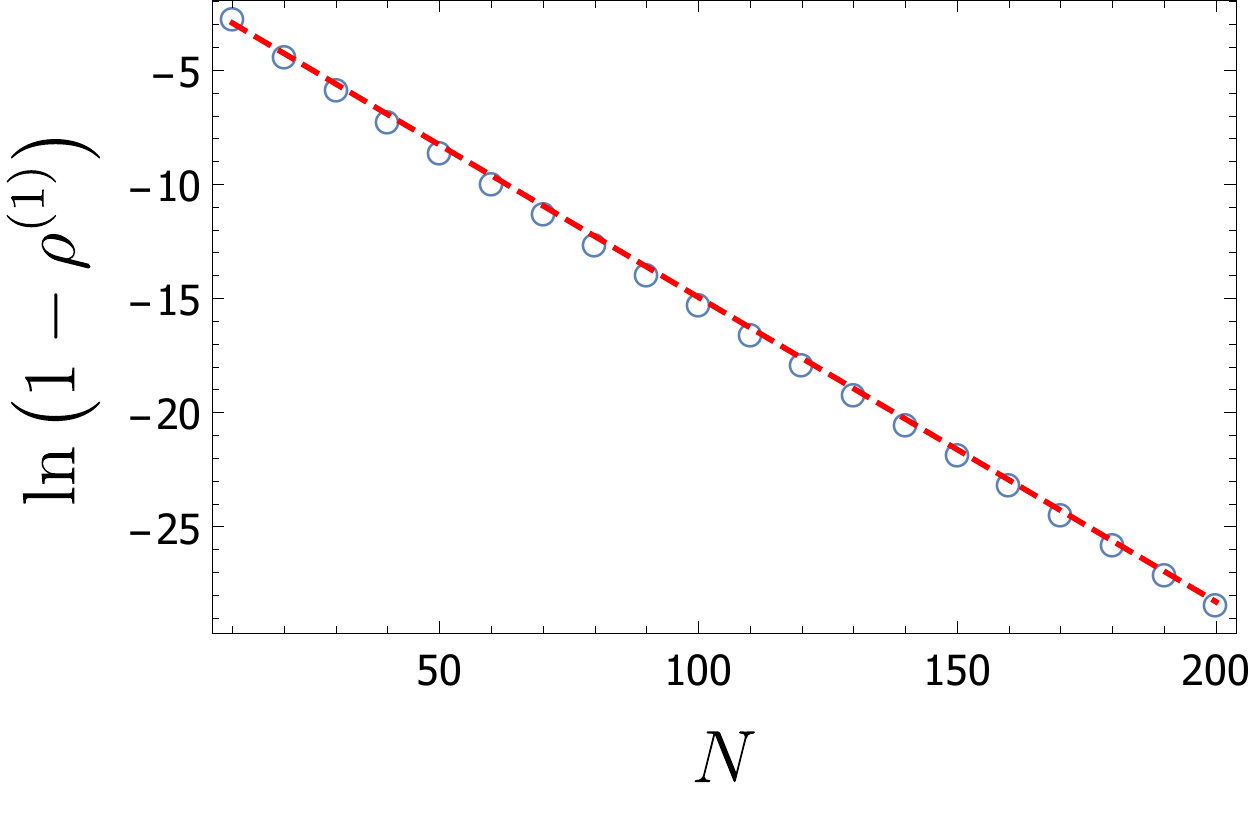}\\
\ \\
\includegraphics[scale=.6]{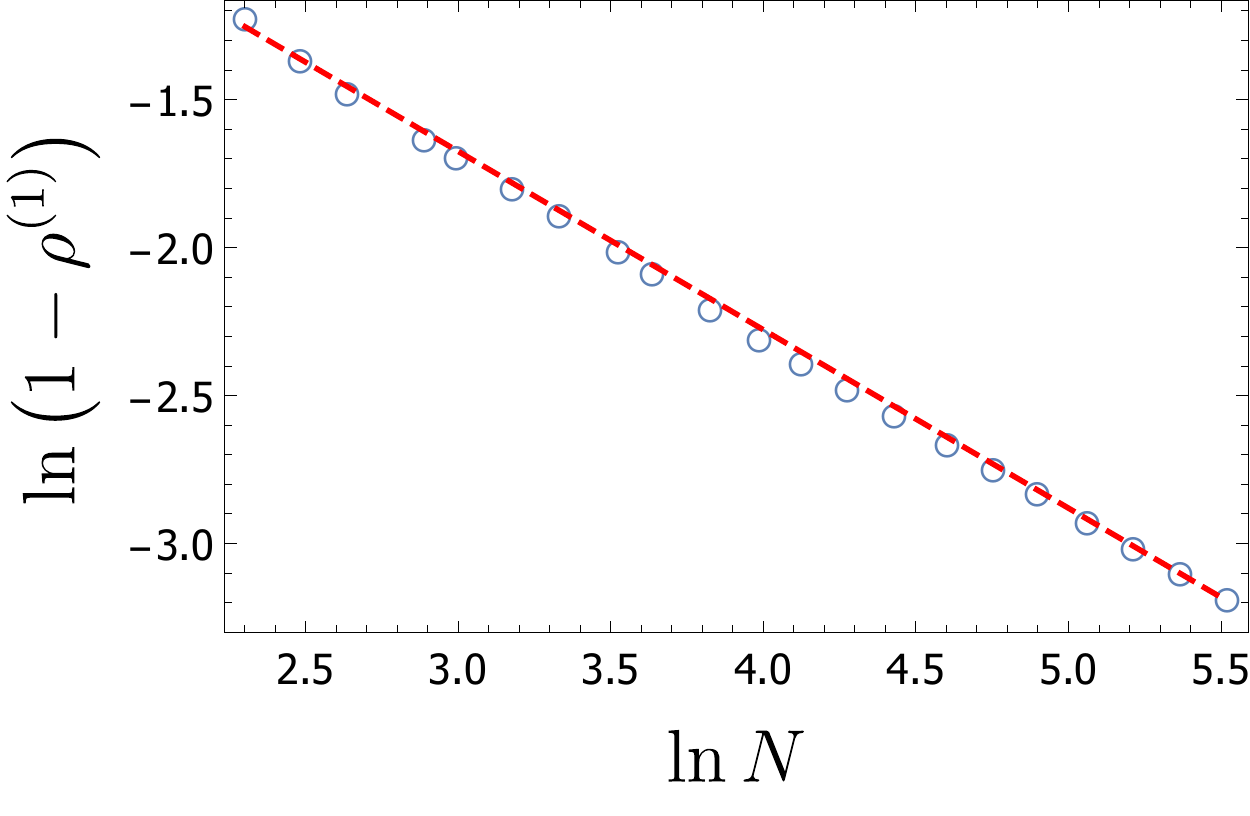}
\caption{Dependence of the largest eigenvalue of the transition matrix $\hat{\bm {Q}}$ with respect to $N$ for the logistic map with $\lambda=3.5$ (top) and $\lambda=4$ (bottom). The red, dashed lines correspond to Eq.~\eqref{eqn:rhostar} with $\alpha\approx 0.2016$, $\beta\approx 0.1334$ (top) and $\gamma\approx 1.1438$, $\delta\approx 0.603$ (bottom). The top panel is representative of the general behavior observed for various choices of $\lambda<4$, as stated in Eq. \eqref{eqn:rhostar}.} 
\label{fig:rhostar}
\end{figure}

The main focus of this paper is, however, the mesoscopic version of the model, which is a good approximation for larger values of $N$ for which the system settles around some stable state and remains there for a very long time, before being eventually absorbed at the boundary. In applications it is frequently this state that we are interested in, not the stationary state. It is commonly referred to as the quasistationary state, since to all intents and purposes it resembles a stationary state of the system, except that if one waits a sufficiently long time the system will `decay' to the $n=0$ state. We would expect, therefore, that it is the second largest eigenvalue of $\bm{Q}$ that is of real interest to us, with $\rho^{(0)}=1$ and $\rho^{(1)} = 1 - \mathcal{O}(e^{-\beta N})$, for some constant $\beta$. We envisage this to be the case, since the decay to the extinction state is via `rare events' which will have the exponentially small character typically found in large deviation theory. This can be tested numerically, where it is indeed found that when $N$ is larger than about $100$, the second eigenvalue is visually indistinguishable from $1$. A more systematic analysis shows the following behavior:
\begin{equation}
\rho^{(1)} \sim \left\lbrace \begin{array}{lcl}
1-\alpha e^{-\beta N} & \ & \lambda < 4\\
1-\gamma N^{-\delta} & \ & \lambda = 4
\end{array}\right. . 
\label{eqn:rhostar}
\end{equation}
Figure \ref{fig:rhostar} shows $\rho^{(1)}$ as a function of $N$ for $\lambda=3.5$ and $\lambda=4$, obtained numerically using the built-in \textsc{Mathematica} function~\cite{Wolfram}. Although not of primary concern here, we note that in the ``critical'' case $\lambda = 4$, the exponential decay is replaced by a power law.


The absorbing state, $n=0$, and the ``true'' stationary state, $P^{\rm st}_n = \delta_{n,0}$, can be eliminated by the following method \cite{darroch_65}. A matrix $\hat{\bm{Q}}$ is constructed by removing the first row and first column of the transition matrix $\bm{Q}$. That is, we write the decomposition of $\bm{Q}$ as
\begin{equation}
\label{Qhat_defn}
\bm{Q}=\left(\begin{array}{rr}
1 \ & \ \bm{r}^{\rm T}_0 \\ \\
\bm{0} \ & \ \hat{\bm{Q}}
\end{array}\right),
\end{equation}
where $\bm{0}$ and $\bm{r}_0$ are vectors of length $N$ and $\hat{\bm{Q}}$ is an $N\times N$ matrix. It is straightforward to show that $C(\rho) = (1-\rho) \hat{C}(\rho)$, where $C(\rho)$ and $\hat{C}(\rho)$ are the characteristic polynomials of $\bm{Q}$ and $\hat{\bm{Q}}$, respectively. From this it follows that (i) the eigenvalues of $\hat{\bm{Q}}$ are those of $\bm{Q}$, apart from the additional one equal to unity, (ii) the left eigenvectors of $\bm{Q}$, other than $(1\,1\,\ldots\,1)^{\rm T}$, have $0$ as their first entry and the left eigenvectors of $\hat{\bm{Q}}$ as the other $N$ entries, and (iii) the right eigenvectors of $\bm{Q}$, other than $(1\,0\,\ldots\,0)^{\rm T}$, have a first entry we denote as $v^{(i)}_0$ and the right eigenvectors of $\hat{\bm{Q}}$ as the other $N$ entries. By orthogonality with $(1\,1\,\ldots\,1)^{\rm T}$, $v^{(i)}_0 = - \sum^{N}_{n=1} v^{(i)}_n$, $i=1,\ldots,N$. 

In summary, the introduction of the matrix $\hat{\bm{Q}}$ means that the absorbing state $n=0$ has been moved outside of the system and is reduced to a collection point for the probability exiting the states $n=1,\ldots,N$. The Perron-Frobenius theorem for non-negative matrices~\cite{meyer_00} can now be applied to $\hat{\bm{Q}}$ to show that the right eigenvector corresponding to its largest eigenvalue $\rho^{(1)}$ has non-negative entries. This is the desired quasistationary distribution, which we will call $\bm{P^{\text{qst}}}$. It is important to note that $\hat{\bm{Q}}$ is not a stochastic matrix: the vector $\bm{r}_0$ has at least one entry not equal to $0$, and so the columns of $\hat{\bm{Q}}$ do not all sum to $1$. As a consequence, the largest eigenvalue of $\hat{\bm{Q}}$, $\rho^{(1)} \neq 1$. Nonetheless, the Perron-Frobenius theorem is still applicable since, despite being widely applied for the case of stochastic matrices, it holds for non-negative matrices in general and does not impose any restriction on the magnitude of the largest eigenvalue~\cite{meyer_00}.

\begin{figure}
\centering
\includegraphics[scale=.68]{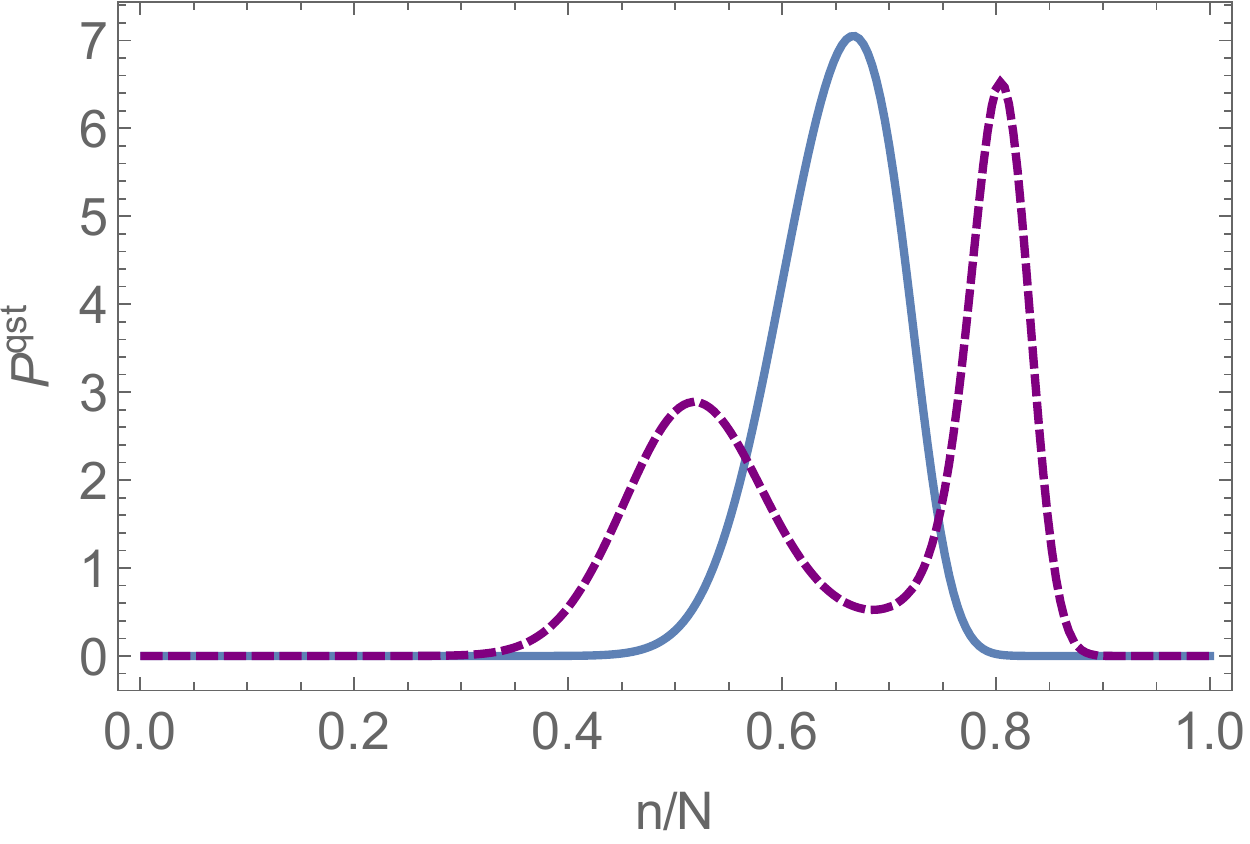}
\caption{Quasistationary distribution of the Markov chain corresponding to the logistic map, obtained as the right eigenvector $\bm v^{(1)}$ associated to the largest eigenvalue of the constrained matrix $\hat{\bm Q}$. Results are shown for both $\lambda=2.9$ (blue solid line) and $\lambda=3.25$ (purple dashed line). In the former case, the logistic map has a stable fixed point; in the latter a stable two-cycle. In both cases $N=250$.}
\label{fig:qsd}
\end{figure}

We may now examine the form of $\bm{P^{\text{qst}}}$ for various values of $\lambda$, the parameter which appears in the logistic map. For $1 < \lambda < 3$, where the logistic map has a single stable fixed point at $z=z^* \neq 0$, it takes the form of a Gaussian centered at the fixed point. As $\lambda$ increases, and passes through $3$, a two-peaked distribution smoothly develops, reflecting the fact that for $3 < \lambda < 1 + \sqrt{6} = 3.449\ldots$ a two-cycle is the stable attractor. The profile of $\bm{P^{\text{qst}}}$ for two parameter choices is shown in  Fig.~\ref{fig:qsd}. Obtaining the eigenvalues and eigenvectors of $\hat{\bm{Q}}$ numerically is not feasible for values of $N$ greater than a few hundred, but the results obtained for $\bm{P^{\text{qst}}}$ for values of $N$ which can be investigated will allow us to verify the output of the mesoscopic version of the model that will be discussed in the following sections.


While it is not possible to explore very large values of $N$ numerically using the Markov chain, some analytic results are available in the infinite $N$ (deterministic) limit, and we will end this section with a discussion of these. In this limit the dynamics is that of the logistic map $z_{t+1}=\lambda z_t(1 - z_t)$. However, we can also formulate this in terms of the distribution of densities found from the kernel of the transfer operator, or Ruelle operator, defined by~\cite{ott_93,ChaosBook}
\begin{equation}
\mathcal{L}^{(t)}(z,w) = \delta\left( z - p^{(t)}(w) \right),
\label{kernel_op}
\end{equation}
which takes the distribution $P_t(w)$ and returns the distribution $P_{t+1}(z)$ one time-step later. Therefore
\begin{eqnarray}
P_{t+1}(z) &=& \mathcal{L}P_t(z) = \int dw\,\delta\left( z - p^{(t)}(w) \right)
P_t(w) \nonumber \\
&=& \sum_{w\in p^{-1}(z)}\frac{P_t(w)}{\vert {\text{det}}J(w) \vert },
\label{transfer_op}
\end{eqnarray}
where $J(w)= p^\prime (w)$ is the Jacobian of the map. We would expect that, in the case of the logistic map, the Gaussian and two-peaked distributions in Fig.~\ref{fig:qsd} would be replaced by their respective $N \to \infty$ limits, namely one delta-function spike located at the fixed point and two delta-function spikes located at the two cycle. That this is so is easily checked. For example, in the case of the fixed point,
\begin{eqnarray}
\int dw\,\mathcal{L}(z,w)\,\delta(w-z^*) &=& \int dw\,\delta(z-p(w))\delta(w-z^*)\nonumber \\
&=& \delta(z-p(z^*)) = \delta(z-z^*),
\label{PF_eigen}
\end{eqnarray}
which shows that $\delta(z-z^*)$ is an eigenfunction of $\mathcal{L}$ with eigenvalue $1$. Similarly one can show that $[\delta(z-z_{-}) + \delta(z-z_{+})]$ is an eigenfunction of $\mathcal{L}$ when a two-cycle at $z_{-}$ and $z_{+}$ exists, also with an eigenvalue $1$. There is no ``escape'' from these states in the $N \to \infty$ limit, and so these are absorbing states. As $\lambda$ increases, so does the number of absorbing states, and so the number of eigenvalues equal to $1$. In dynamical systems theory the quasistationary distribution in the $N \to \infty$ limit is referred to as the invariant measure, which in the chaotic regime can have a very complex structure, as opposed to the delta-function spikes we have seen for the fixed point and two-cycle. When $\lambda = 4$ the invariant density for the logistic map can be found analytically \cite{strogatz_94} and is given by $[\pi \sqrt{z(1-z)} ]^{-1}$.

As we have already stressed, the microscopic model, which takes the form of a finite Markov chain, is difficult to analyze and can only be numerically investigated for relatively small values of $N$. To make further progress we construct a mesoscopic \cite{gardiner_09} version of the model, in which we lose some of the finer details, but still retain enough of the structure of the model so as to allow accurate predictions at moderate to large values of $N$. 

\section{Mesoscopic formulation}
\label{sec:meso}
The derivation of a mesoscopic model from a microscopic one, where the discrete lattice points are replaced by a continuous variable $z=n/N$, is quite standard in the theory of continuous-time Markov processes \cite{gardiner_09,risken_89}, and the derivation for the discrete-time case initially follows the same route. That is, one begins from the Chapman-Kolmogorov equation---which holds because the process is Markovian---and rewrites it so it takes the form of a Kramers-Moyal expansion \cite{gardiner_09,risken_89}:
\begin{equation}
P_{t+1}(z) = \sum_{\ell=0}^{\infty} \frac{(-1)^{\ell}}{\ell!} 
\frac{\partial^{\ell} }{\partial z^{\ell}} \left[ M_{\ell}(z) P_t(z) \right],
\label{KM_conventional}
\end{equation}
where $M_{\ell}(z)$ is the $\ell$th jump moment and is defined by
\begin{equation}
M_{\ell}(z) = \langle \left[z_{t+1} - z_{t}\right]^{\ell} \rangle_{z_{t}=z}.
\label{defn_M}
\end{equation}
Here we have suppressed the dependence on the initial conditions in both $P_{t+1}(z)$ and $P_t(z)$. The jump moments only involve one time step and can be calculated for the particular microscopic model one wishes to study. It is frequently found that that for $\ell>2$ they are small, and the infinite series can be truncated at $\ell=2$. 

Here, however, this is not the case. If time is continuous, $z_{t+1}$ appears in Eq.~(\ref{defn_M}) as $z_{t+\delta t}$ and is infinitesimally close to $z_t$, whereas in the discrete-time case large jumps are allowed. Furthermore, the distribution of values for $z_{t+1}$ will not in general be centered around $z_t$. Progress can be made by noting that $z_{t+1}$ is expected to be close to $p_t \equiv p(z_t)$ for large $N$, since they are equal in the macroscopic limit, this being the definition of the equivalent deterministic map. We can now rewrite the jump moments as \cite{challenger_14} 
\begin{eqnarray}
M_{\ell}(z) &=& \langle \left[z_{t+1} - p_t + 
\left\{ p_t - z_{t} \right\} \right]^{\ell} \rangle_{z_{t}=z} \nonumber \\
&=& \langle \left[z_{t+1} - p_t + 
\left\{ p - z \right\} \right]^{\ell} \rangle_{z_{t}=z},
\label{new_M}
\end{eqnarray}
where $p=p(z_t)|_{z_{t}=z}$. Expanding out Eq.~(\ref{new_M}) gives
\begin{equation}
M_{\ell}(z)
= \sum^{\ell}_{r=0}\, {\ell \choose r} 
\left( p - z \right)^{\ell - r} J_r(p),
\label{M_to_J}
\end{equation}
where
\begin{equation}
J_r(p) \equiv \left\langle \left[ z_{t+1} - p_t \right]^r \right\rangle_{z_{t}=z}.
\label{J_r}
\end{equation}
As expected, these jump moments do fall off with $N$: for $r>2$ $J_r(p)=\mathcal{O}(N^{-2})$ \cite{challenger_14}, and so a truncation can be effected by neglecting the $J_r(p)$ for $r > 2$. This approximation applied to Eq.~(\ref{KM_conventional}) leads to 
\begin{eqnarray}
& & P_{t+1}(z) = \sum_{\ell=0}^{\infty} \frac{(-1)^{\ell}}{\ell!} 
\frac{\partial^{\ell}}{\partial z^{\ell}} \left[ (p-z)^{\ell}\,P_t(z)\right]
\nonumber \\ & & +\frac{1}{2N}\,\sum^{\infty}_{\ell = 0}\,\frac{(-1)^{\ell}}{\ell !}\,
\frac{\partial^{\ell}}{\partial z^{\ell}}\,\frac{\partial^2}{\partial z^2}\,
\left[ (p-z)^{\ell}\,p(1-p)\,P_t(z)\right] \nonumber \\ & & + 
\mathcal{O}\left(\frac{1}{N^2}\right).
\label{strange_KM_expansion}
\end{eqnarray}
In the continuous-time case this truncation would lead to a Fokker-Planck equation---a partial differential equation which is second order in the $z$ variable and first order in time. Here we find an equation which is infinite order in the $z$ variable and a difference equation in time. However, in Ref.~\cite{challenger_14} we showed that a change of variables, allied to the use of the Fourier transform, allows us to write the equation as
\begin{equation}
P_{t+1}(z) = {\cal P}_t(z) + \frac{1}{2N}\,\frac{\partial^2 }{\partial z^2} 
\left[ z \left( 1 - z \right) \mathcal{P}_t(z) \right] +
\mathcal{O}\left(\frac{1}{N^2}\right). 
\label{second_order_form}
\end{equation}
The pdf $\mathcal{P}$ enters through a change of variables and is defined by $P_t(z)dz=\mathcal{P}_t(p)dp$. Note that we can write the $N\to \infty$ limit of the equation above explicitly in terms of $P$ as Eq.~\eqref{transfer_op}. Equation \eqref{second_order_form}, for finite $N$, is the discrete-time equivalent of a Fokker-Planck equation. 

Our interest in this paper is intimately connected with the structure of Eq.~(\ref{strange_KM_expansion}). Since it has derivatives of infinite order, presumably an infinite number of boundary conditions need to be specified. Thus we are led to conclude that the entire functional form of $P_t(z)$ needs to be specified for $z < 0$ and $z > 1$ for all $t$. This would fit in with our intuition, that finite jumps in this mesoscopic formulation can now take us out of the interval $0 \leq z \leq 1$---which was not the case in the microscopic formulation --- and need to be forbidden by extra conditions. In fact, in the derivation of Eq.~(\ref{second_order_form}) \cite{challenger_14} we had to explicitly make the assumption that $P_t(z) \equiv 0$ for $z < 0$ and $z > 1$ for all $t$, and this clearly is the additional data that needs to be specified in addition to either Eq.~(\ref{strange_KM_expansion}) or Eq.~(\ref{second_order_form}).

In continuous-time processes, working with stochastic differential equations, rather than with the equivalent Fokker-Planck equation, is frequently more intuitive, and this is even more so in the discrete-time case. The equation is now a stochastic \textit{difference} equation (SDE) and takes the form \cite{challenger_14}
\begin{equation}
z_{t+1} = p_t + \eta_t,
\label{stoch_diff_eqn}
\end{equation}
where $\eta_t$ is a Gaussian variable with zero mean and correlator
\begin{equation}
\langle \eta_t \eta_{t'} \rangle = \frac{p(1-p)}{N} \delta_{t\,t'}.
\label{correlator}
\end{equation}
We have argued above that the differential-difference equation for $P_t(z)$ should be augmented with the condition that it vanishes outside of the interval $0 \leq z \leq 1$. In terms of Eq.~(\ref{stoch_diff_eqn}) this means that trajectories should not be allowed to leave the interval. But how can this rule be implemented in practice?

It should be stressed that this question has little connection with ones relating to conventional boundary conditions. For instance, in the microscopic formulation of Sec.~\ref{sec:micro}, only conditions at the actual boundary need be considered. These in any case are determined naturally from the nature of the transition matrix $\bm{Q}$: an absorbing boundary at $n=0$ and a transition from $n=N$ to $n=0$ with probability $1$. Similarly, in the analogous stochastic differential equation, where time is continuous, the multiplicative form of the noise would resolve this problem since the noise vanishes at the boundary. However, because the system we are studying here admits large jumps, it is possible to jump directly from inside to outside of the interval. So although the noise vanishes at the boundary, this is not sufficient to prevent trajectories escaping. 

It is possible to postulate reasonable rules to deal with this ``leakage'' of probability out of the system. In simulations of the SDE one could simply discard trajectories that at some time exited the interval $(0,1)$ and restart the simulation. Or one could restart such trajectories in some random position $z \in [0,1]$. The former method is how we proceeded in previous work where we compared exact results from the Markov chain in Eq.~(\ref{markov_chain}) with numerical simulation results from the stochastic difference equation and found good agreement~\cite{challenger_13,challenger_14}. However we believe this question needs a more systematic analysis, especially when $N$ is not large and also when $\lambda$ takes a value close to 4, where the invariant density of the deterministic map covers most of the unit interval (see Sec.~\ref{sec:micro}).


\begin{figure}[h!]
\centering
\subfigure{\includegraphics[scale=.6]{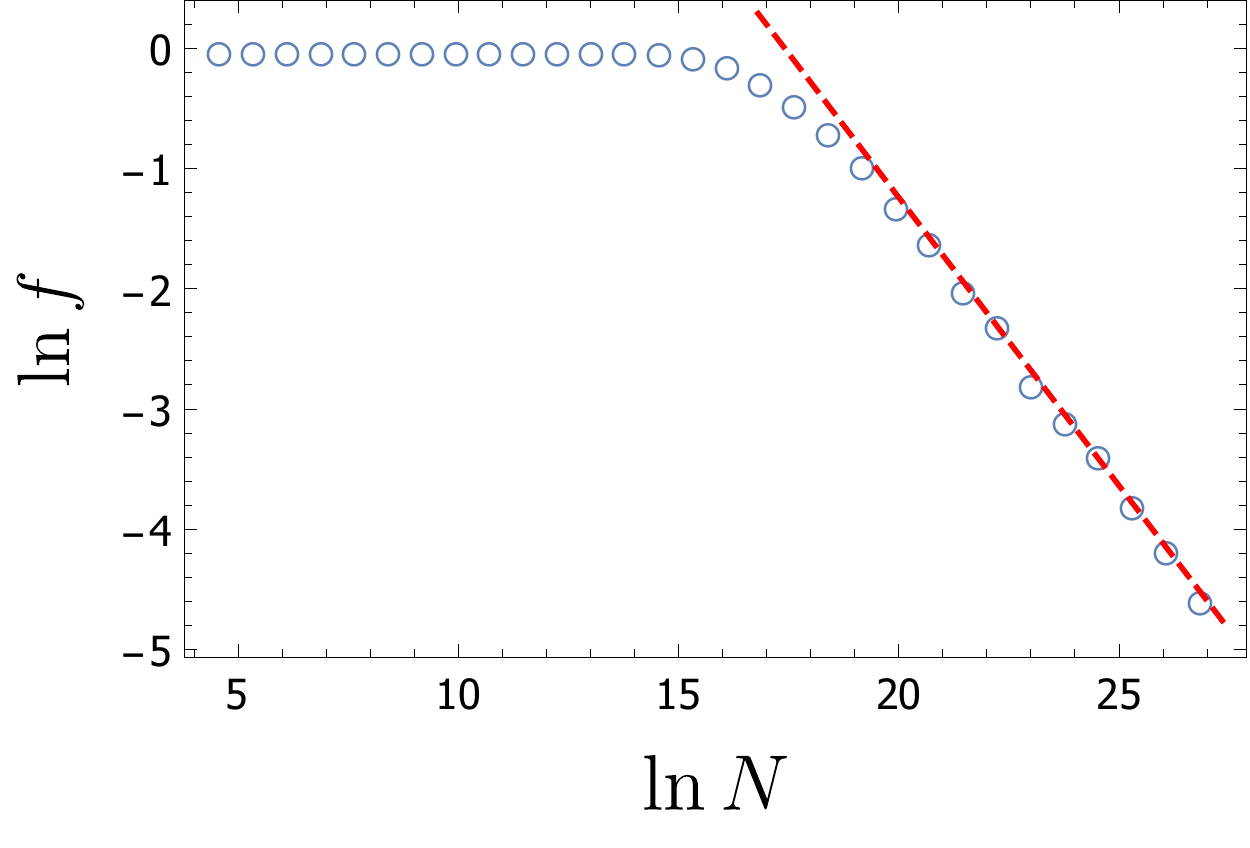}}\\
\subfigure{\includegraphics[scale=.6]{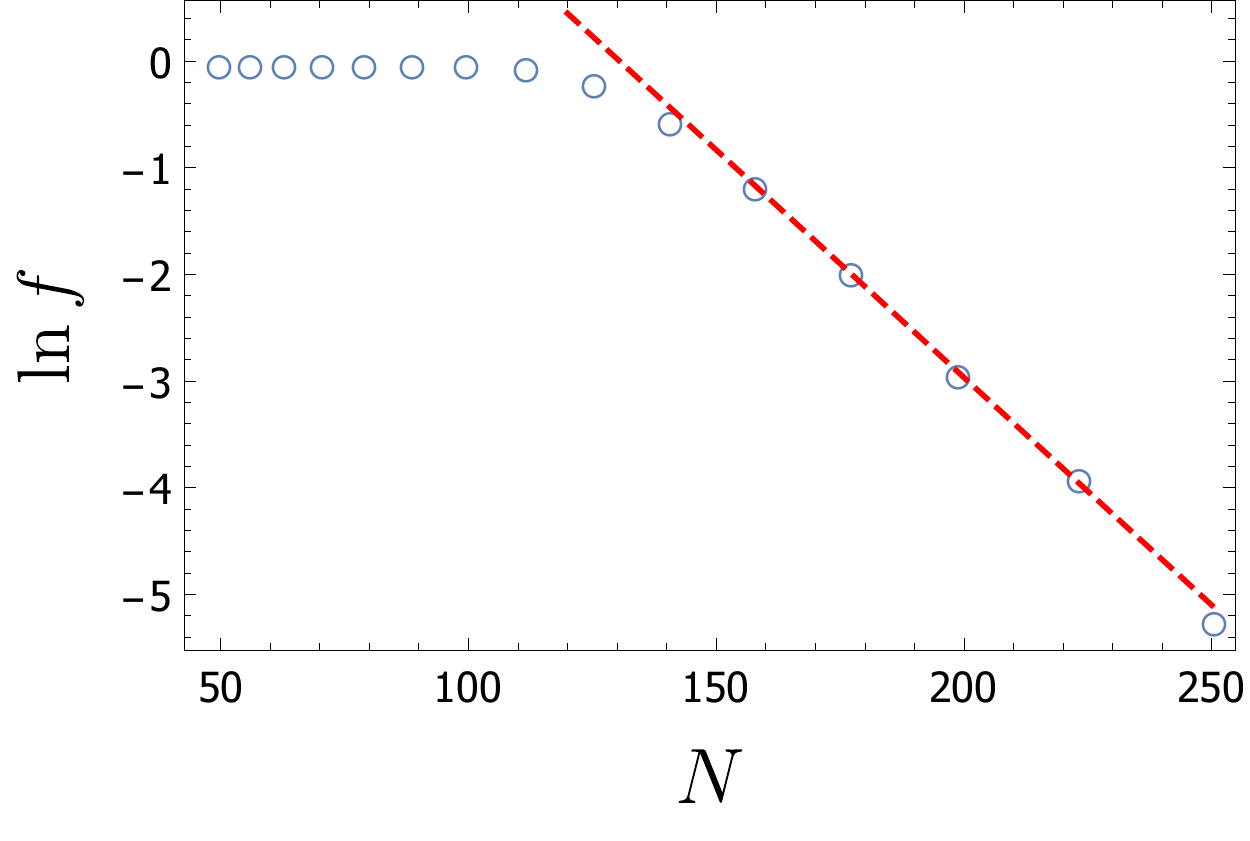}}
\caption{Fraction of trajectories, out of 5000, that leave the unit interval over the course of 20000 iterates, as a function of $N$ for the logistic map with $\lambda=4$ (top) and $\lambda=3.7$ (bottom). The red dashed lines correspond to $f\approx 4235.25 N^{-0.48}$ (top) and $f\approx 261.7 e^{-0.0427N}$ (bottom).}
\label{fig:LeakFraction}
\end{figure}

\begin{figure}
\centering
\subfigure{\includegraphics[scale=.6]{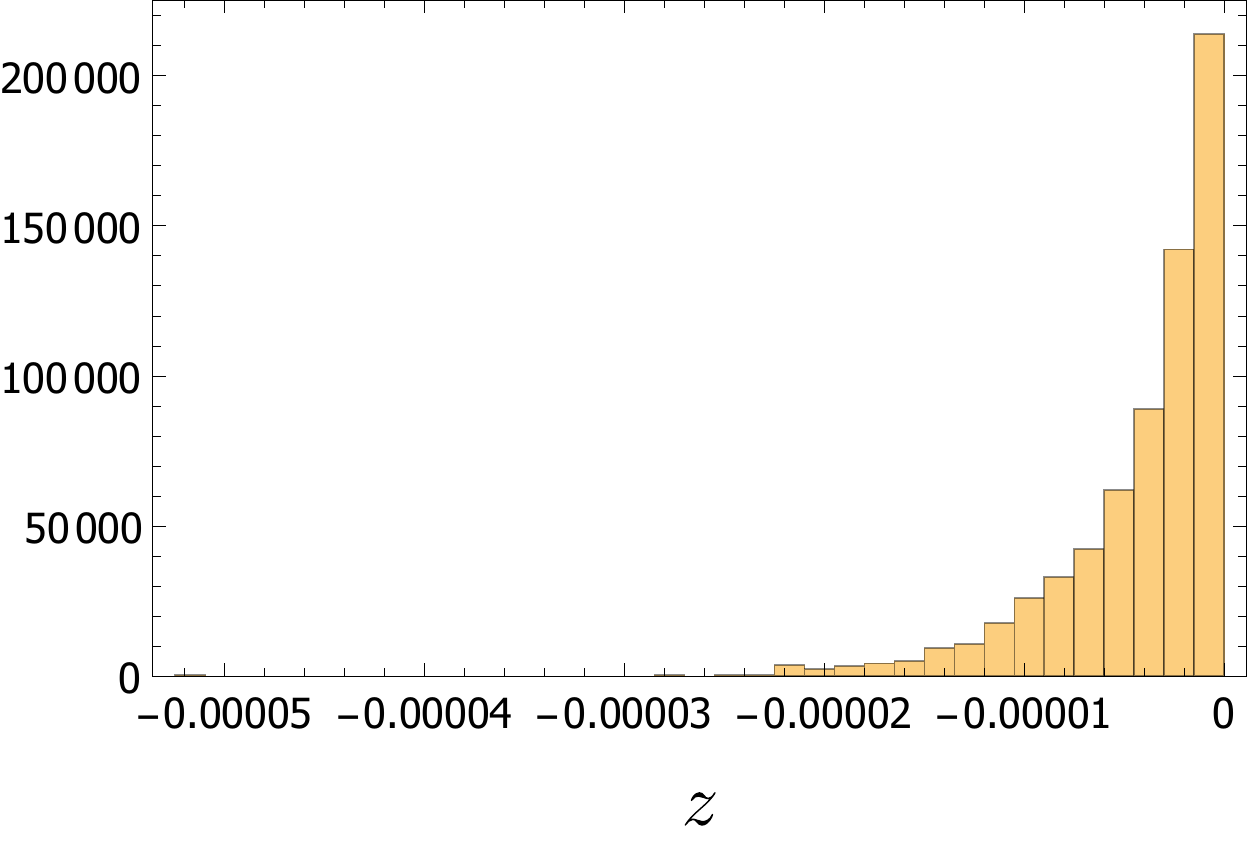}}\\
\subfigure{\includegraphics[scale=.6]{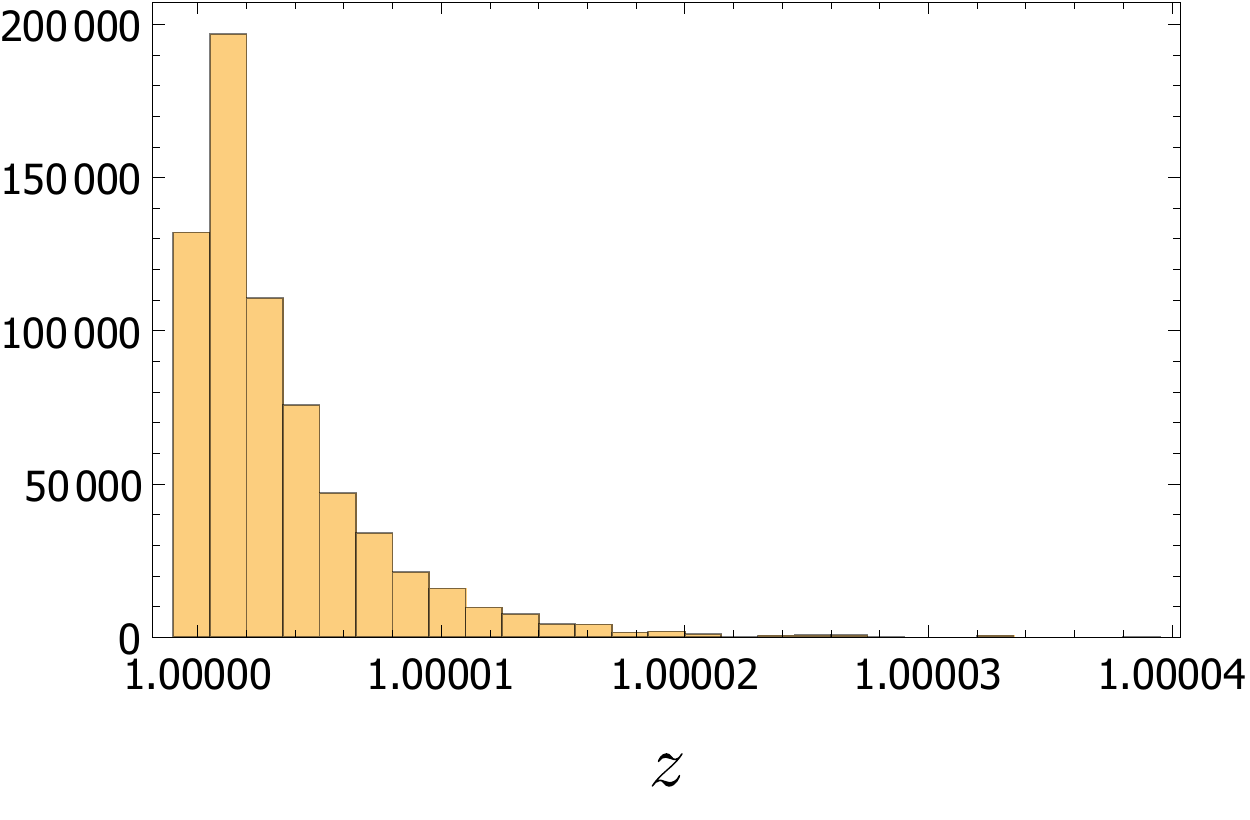}}
\caption{Distribution of first iterates leaving the unit interval from the boundaries at $z=0$ (top) and $z=1$ (bottom) for the logistic map with $\lambda=4$, $N=10^5$.}
\label{fig:LeakDist}
\end{figure}

Before devising algorithms to deal with this leakage in Sec.~\ref{sec:trun_cen}, we need to have some idea of how often it happens. We carried out numerical simulations of the SDE (\ref{stoch_diff_eqn}) for $\lambda = 4$, since as explained above we expect the effect to be most noticeable here. The top panel of Fig.~\ref{fig:LeakFraction} shows the fraction $f$ of trajectories that leave the unit interval as a function of $N$ in the logistic map with $\lambda=4$, from a total of 5000 realizations of the SDE, of 20000 iterations each. This fraction is found to scale roughly as $N^{-1/2}$. As a comparison, the same quantity is plotted for $\lambda=3.7$ in the bottom panel; the fraction of trajectories leaving the interval falls off exponentially, and escape events are very rare even for relatively low values of $N$. One can also ask where the exits from the interval land. Figure \ref{fig:LeakDist} shows the distribution of first iterates to leave the unit interval, both through the lower and upper boundaries, for the case $\lambda=4$ and $N=10^5$.


One of the most significant aspects of the approximation which takes us from the microscopic model in the form of Eq.~(\ref{markov_chain}) to the SDE (\ref{stoch_diff_eqn}) is the neglect of jump moments $J_r(p)$ for $r > 2$. We end this section by examining the consequence of this by performing the calculation of the higher jump moments (from the original, microscopic model) numerically, to check whether this truncation is reasonable. As shown in Appendix \ref{App_1}, in a mesoscopic context, the jump moments are simply the moments of the noise $\eta_t$ at time $t$. Therefore these can be used to numerically calculate the characteristic function corresponding to the probability distribution of the noise. The probability distribution can then be obtained by taking the Fourier transform of the characteristic function. We denote this distribution by $\Pi(\eta | p)$, since it will depend on the value of $p$. Here we present some results graphically: further details can be found in Appendix \ref{App_1}. 

\begin{figure}
 \begin{center}
  \subfigure{\includegraphics[scale=0.58]{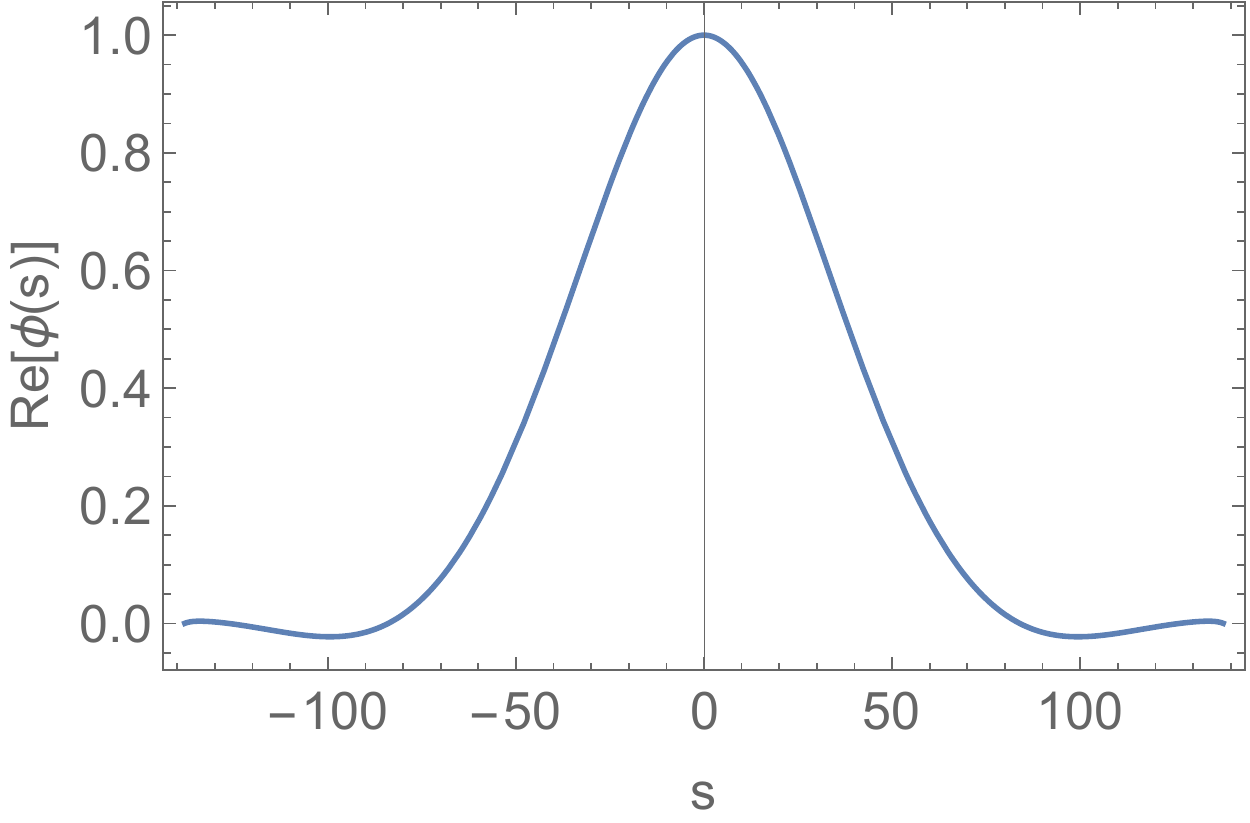}}\\
\subfigure{\includegraphics[scale=0.61]{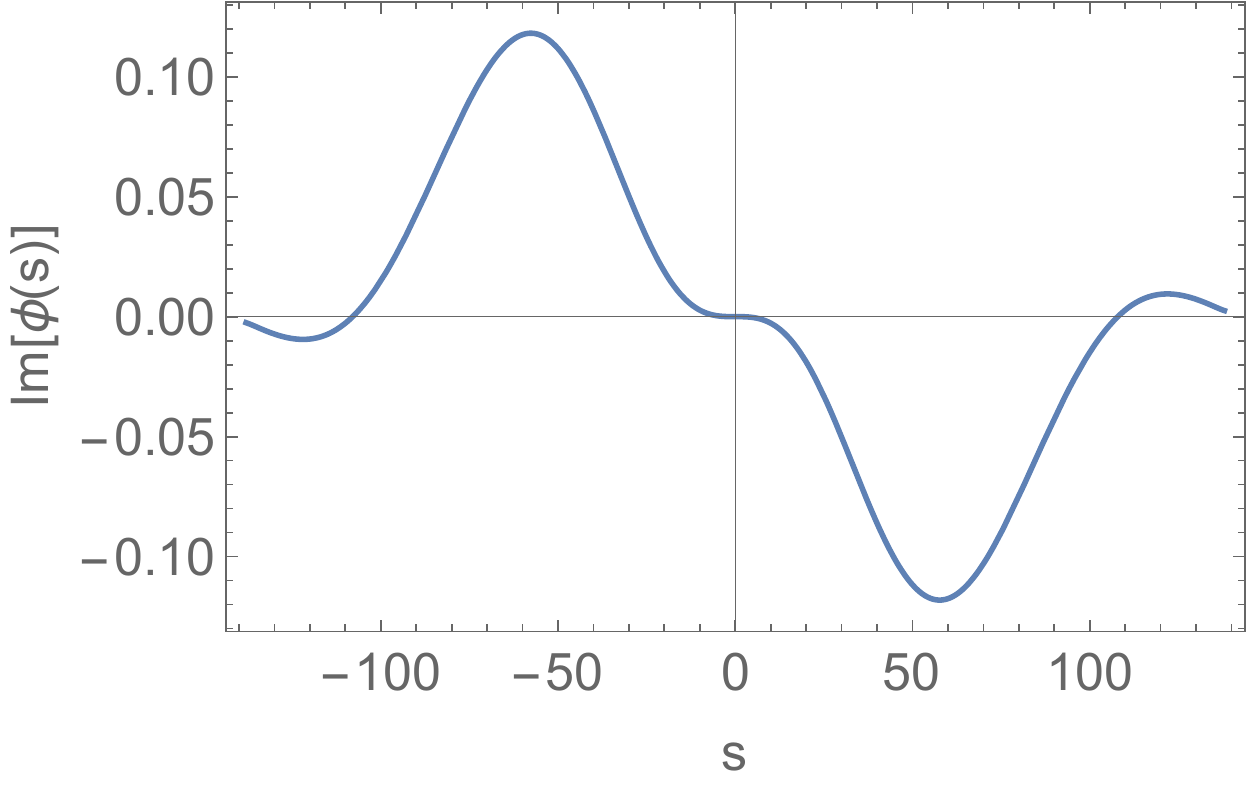}}
 \end{center}
\caption{The real (top) and imaginary (bottom) parts of the characteristic function $\phi(s)$. Parameters values are $p=0.05$ and $N=50$.}
\label{fig:char_func}
\end{figure}

\begin{figure}
 \begin{center}
  \subfigure{\includegraphics[scale=0.65]{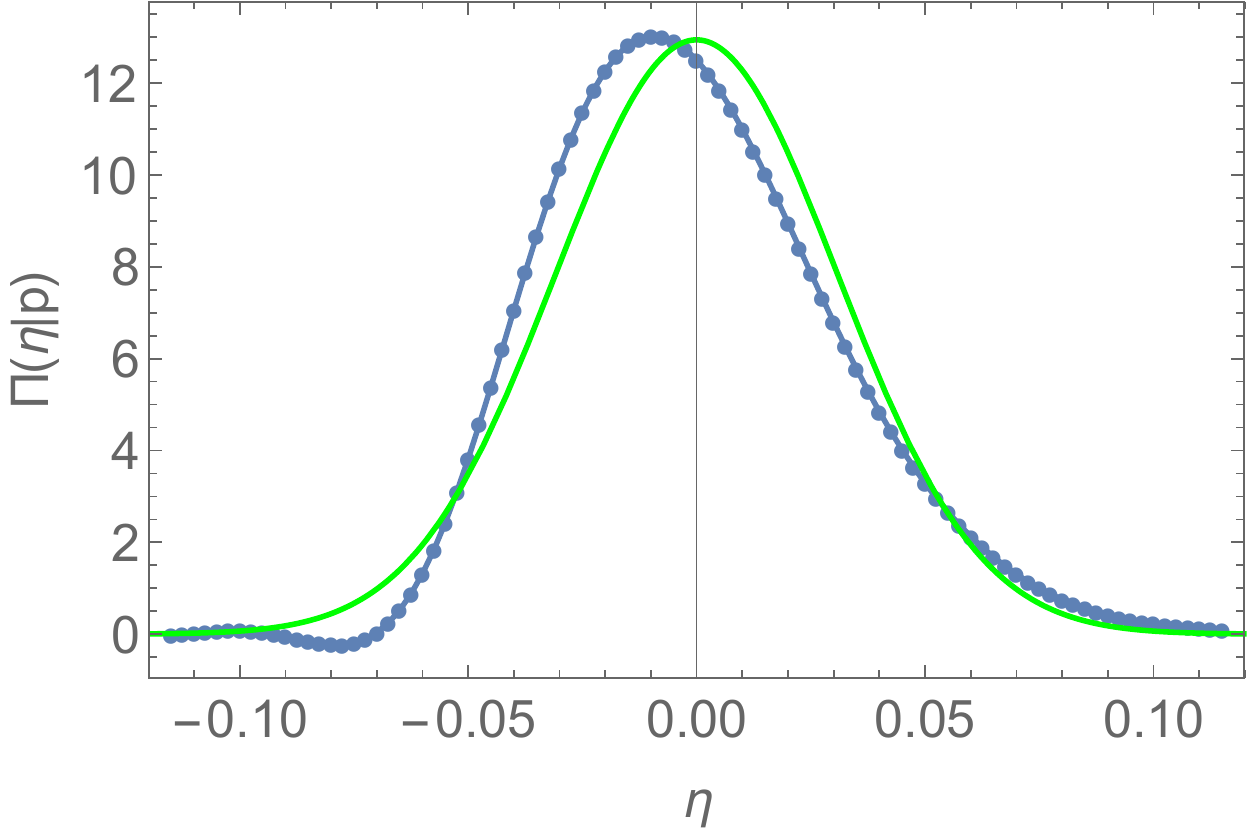}}\\
\subfigure{\includegraphics[scale=0.65]{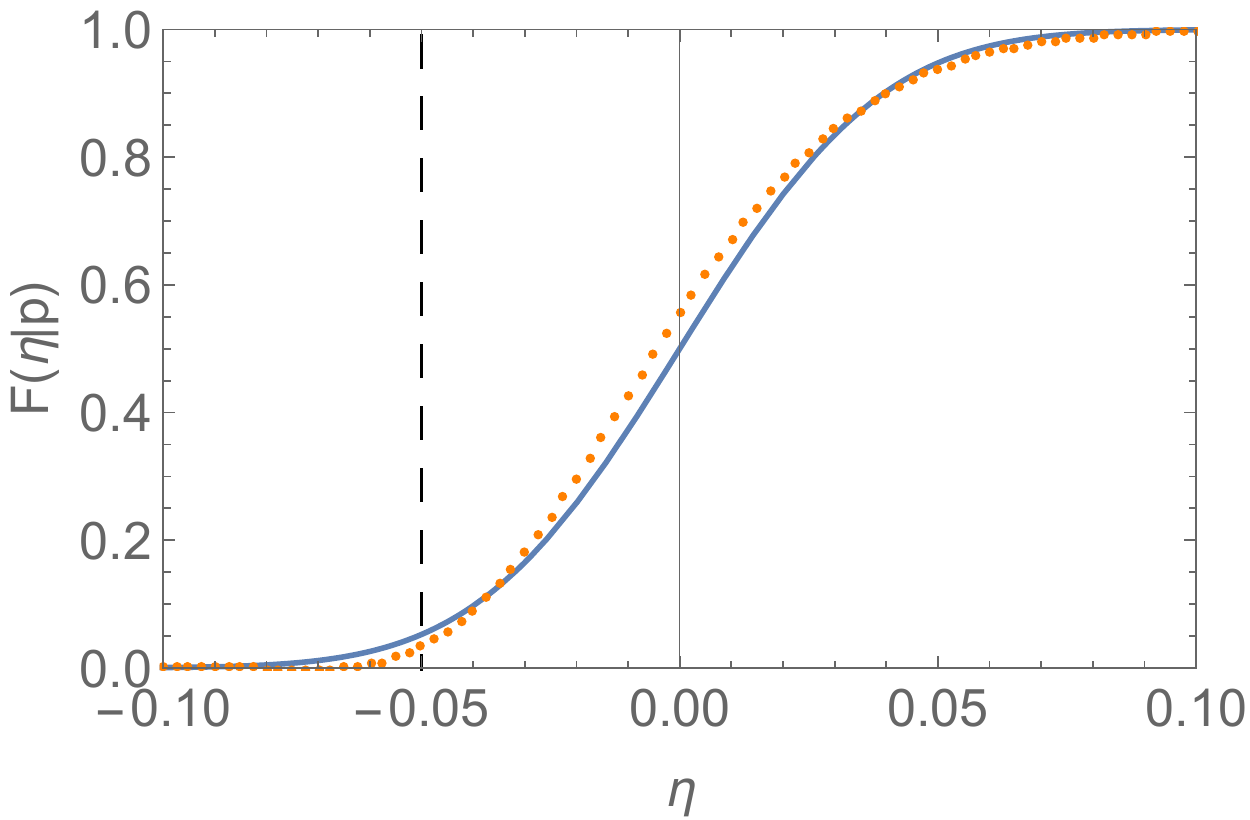}}
 \end{center}
\caption{Top: The probability distribution for the noise, calculated using the higher jump moments (blue dots), compared with the Gaussian noise used in the stochastic difference equation (green solid line). The parameter values are $p=0.05$ and $N=50$. Bottom: the cumulative distributions for the same parameter values. The orange dots indicate the distribution calculated using the higher jump moments, while the cumulative distribution for the Gaussian noise is shown by the blue solid line. The region to the left of the dashed vertical line indicates values of the noise that cause a trajectory to leave the interval.}
\label{fig:prob_dist}
\end{figure}

We would expect non-Gaussian effects to be most noticeable when $N$ is small and $p=p(z)$ is very close to one of the boundaries. Under these conditions it is indeed found that the probability distribution for the noise becomes noticeably skewed and therefore nonsymmetric about the origin. Such a distribution is indicated by a characteristic function that has a nonzero imaginary component, as shown in Fig.~\ref{fig:char_func}~\cite{kolassa_13}. The corresponding probability distribution is shown in Fig.~\ref{fig:prob_dist}, where it is compared with the noise distribution used in the stochastic difference equation given in Eq.~\eqref{stoch_diff_eqn}, which is Gaussian with zero mean and with a variance given by Eq.~(\ref{correlator}). Figure \ref{fig:prob_dist} also shows the two cumulative distributions. Although the consideration of the higher jump moments does introduce a measurable skew to the distribution, so that now less probability lies in the forbidden region, it does not remove escape events from the system, as the lower panel of Fig.~\ref{fig:prob_dist} clearly shows. We also repeated the procedure for larger values of $N$. As the value of $N$ increases, the Gaussian distribution for the noise becomes more and more accurate, and Eq.~\eqref{stoch_diff_eqn} can safely be used to model the microscopic process. Note, however, that very large values of $N$ are required as the deterministic map approaches the boundaries, as illustrated in Fig.~\ref{fig:LeakFraction} for the extreme case $\lambda=4$.

\begin{figure}
\centering
\includegraphics[scale=0.6]{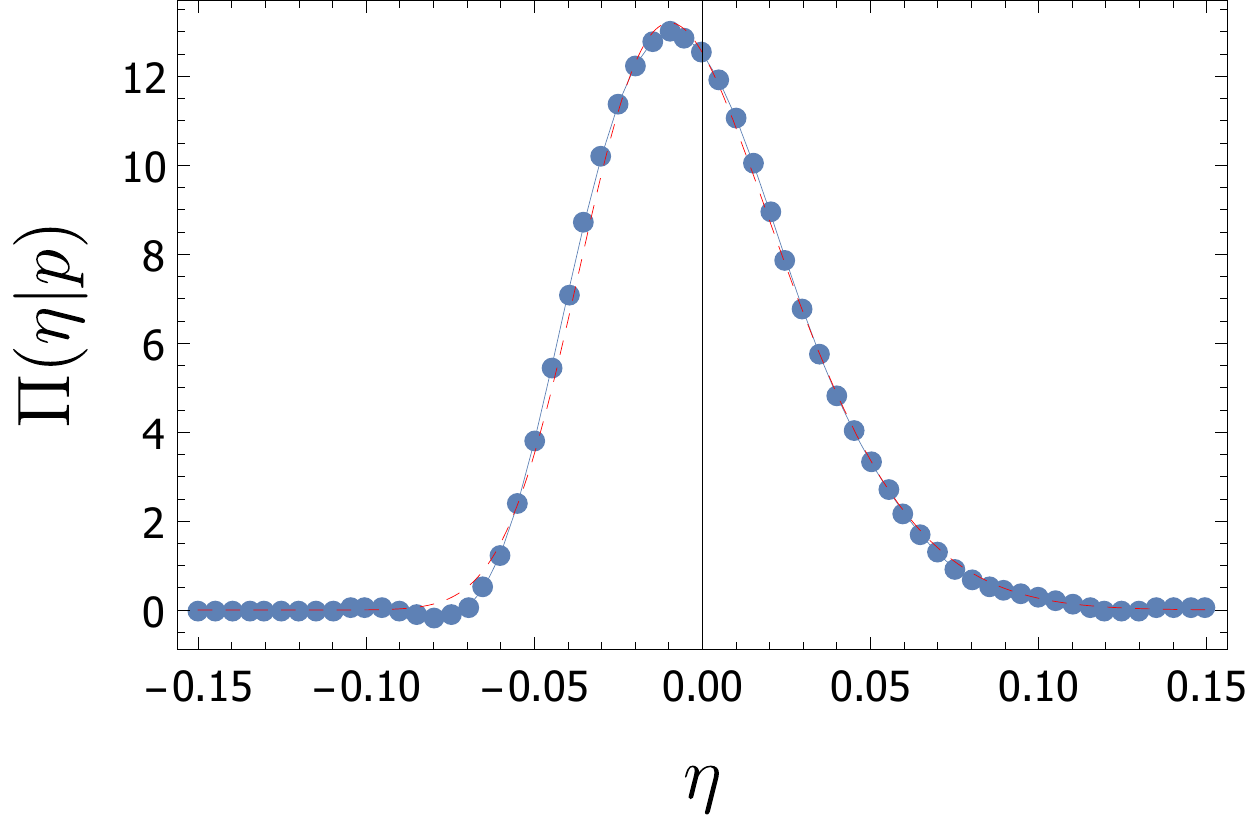}
\caption{The probability distribution for the noise, calculated using the higher jump moments (blue dots), compared with a skew-normal distribution with parameters given by Eq. \eqref{para_skew_dist} (red dashed line). The parameter values are $p=0.05$ and $N=50$.}
\label{fig:skew}
\end{figure}

In order to use such a non-Gaussian distribution in the next section, we need to parametrize it in some way. This is due to the fact that computing the characteristic function of the noise distribution at every iteration of the SDE would be extremely time consuming. Therefore, we choose to fit the noise distribution to a skew-normal distribution with shape parameter $\alpha$, scale parameter $\omega$, and location parameter $\xi$~\cite{azzalini_85} given by
\begin{eqnarray}
\begin{split}
\alpha &= \alpha_0 \beta,\\
\omega &= \omega_0 \left[\frac{p(1-p)}{N}\right]^{1/2} \left[1-\frac{2 \alpha^2}{\pi (1+\alpha^2)}\right]^{-1/2},\\
\xi &= p - \alpha \omega \sqrt{\frac{2}{\pi (1+\alpha^2)}},
\label{para_skew_dist}
\end{split}
\end{eqnarray}
where, away from the boundaries, $\alpha_0= 4$ and $\omega_0= 1$; meanwhile, $\beta = (1-2p)/\sqrt{N p(1-p)}$ is the skewness of the binomial distribution with parameters $N$ and $p$~\cite{abramowitz_65}. We choose this distribution for the simple reason that it is a natural generalization of the normal distribution to allow for nonzero skewness. The choice of parameters ensures that the mean and variance of the distribution are independent of $\alpha$ and are, respectively, $p$ and $\omega_0^2 p(1-p)/N$. Furthermore, for $p\to 1/2$, $\alpha \to 0$ and $\omega_0= 1$, and the distribution tends to a Gaussian distribution with variance $p(1-p)/N$ centered on $p$, which corresponds to the distribution specified by Eq.~\eqref{correlator}. Near the boundaries, we take
\begin{eqnarray}
\begin{split}
\alpha_0 &= 3.2 + p_1\sqrt{3N},\\
\omega_0 &= 1-\dfrac{p_1}{25}+\dfrac{0.05}{\sqrt{p_1 N}},
\end{split}
\end{eqnarray}
with $p_1=\min \lbrace p,1-p \rbrace$. These values are an approximation to the ones obtained by fitting the PDF of a skew-normal distribution, with parameters given by Eq.~\eqref{para_skew_dist}, to the binomial distribution for the noise. Figure~\ref{fig:skew} shows the skew-normal distribution given by~\eqref{para_skew_dist} for the same choice of parameters as in Fig.~\ref{fig:prob_dist}.


\section{The truncated and censored noise distributions}
\label{sec:trun_cen}
Numerical studies of the model expressed as a Markov chain [Eq.~(\ref{markov_chain})], using standard tools such as \textsc{Mathematica}~\cite{Wolfram}, are only feasible up to values of a few hundred for $N$. In the mesoscopic formulation [Eq.~\eqref{stoch_diff_eqn}], on the other hand, $N$ appears as a parameter, and therefore arbitrarily large values can be investigated. The problem, as we have discussed in detail in Sec.~\ref{sec:meso}, is that in the numerical simulation of the SDE (\ref{stoch_diff_eqn}), trajectories which escape outside of the interval $[0,1]$ have to be somehow discounted. Simple rules for implementing this have already been mentioned, for example rejecting escaping trajectories and restarting the simulation or resetting the value of $z$ to an arbitrary point in $(0,1)$. Here we examine two more methodical approaches to this question.

A more sophisticated prescription would be to use a suitably defined noise distribution to make a choice for the noise which ensures that the trajectory does not exit the boundary. In Sec.~\ref{sec:meso} we discussed possible forms of the noise distribution. Here we now ask: what is the microscopic distribution to which this noise distribution is equivalent, and does it have any properties that we would wish to impose on $\Pi(\eta|p)$? This question is discussed in Appendix \ref{App_2}, where we argue that the $m$th column of the transition matrix $\bm Q$ fulfills this role; we recall that this is a binomial distribution with parameters $N$ and $p$. Here $m$ is fixed such that $p=p(m/N)$, and $n$ corresponds to $N(p+\eta)$. Since $Q_{n m}$ vanishes for $n<0$ or $n>N$, this suggests that we should demand that $\Pi(\eta|p)$ is zero for noises such that $\eta < -p$ and $\eta > (1-p)$. Using such a distribution does indeed seem reasonable, since starting at $p$ implies [using Eq.~(\ref{stoch_diff_eqn})] that we move to $z_{t+1}=p+\eta$ in one jump. But if $\eta$ is drawn from a distribution for which $\eta$ can never be smaller than $-p$ or larger than $1-p$, this means that a $z_{t+1}$ is never found which is less than $0$ or greater than $1$, and so the SDE (\ref{stoch_diff_eqn}) can safely be simulated ignoring escape events altogether. 

\begin{figure}[h!]
\centering
\subfigure{\includegraphics[scale=.6]{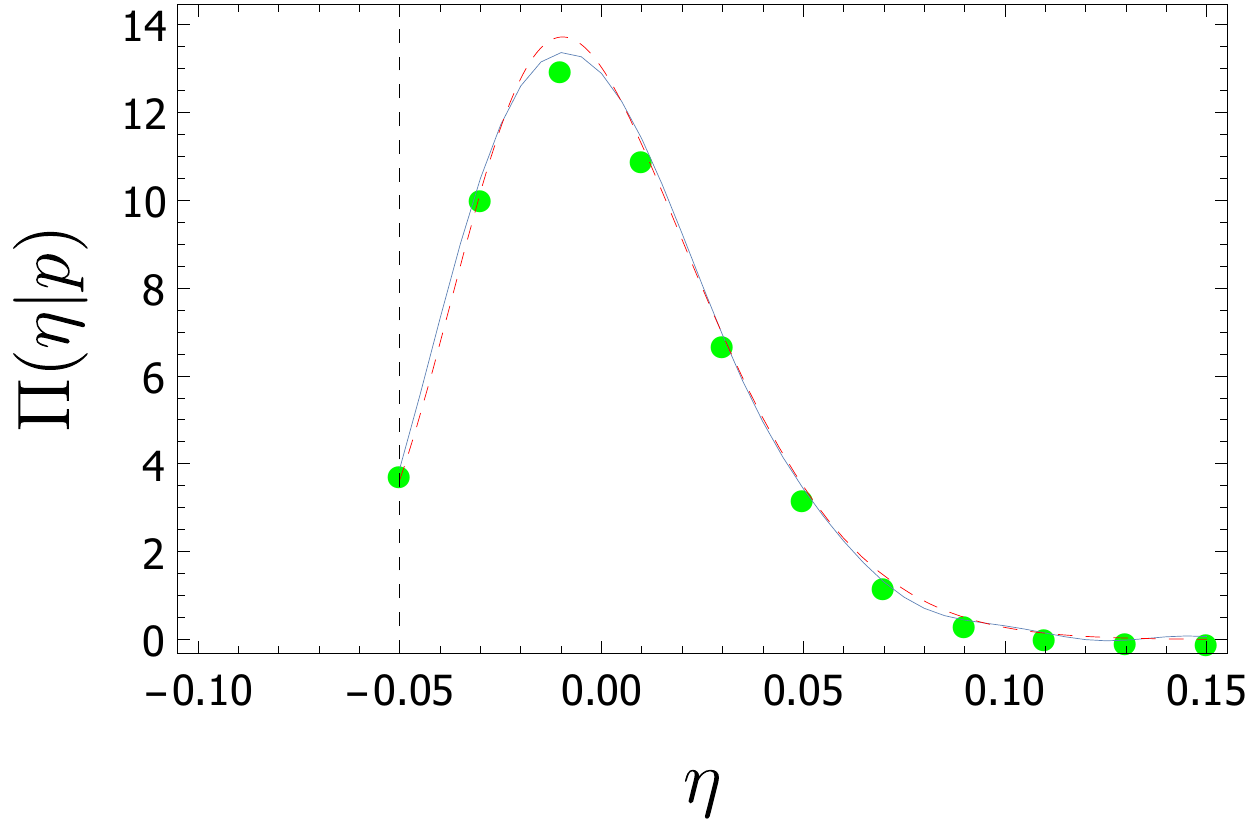}}\\
\subfigure{\includegraphics[scale=.6]{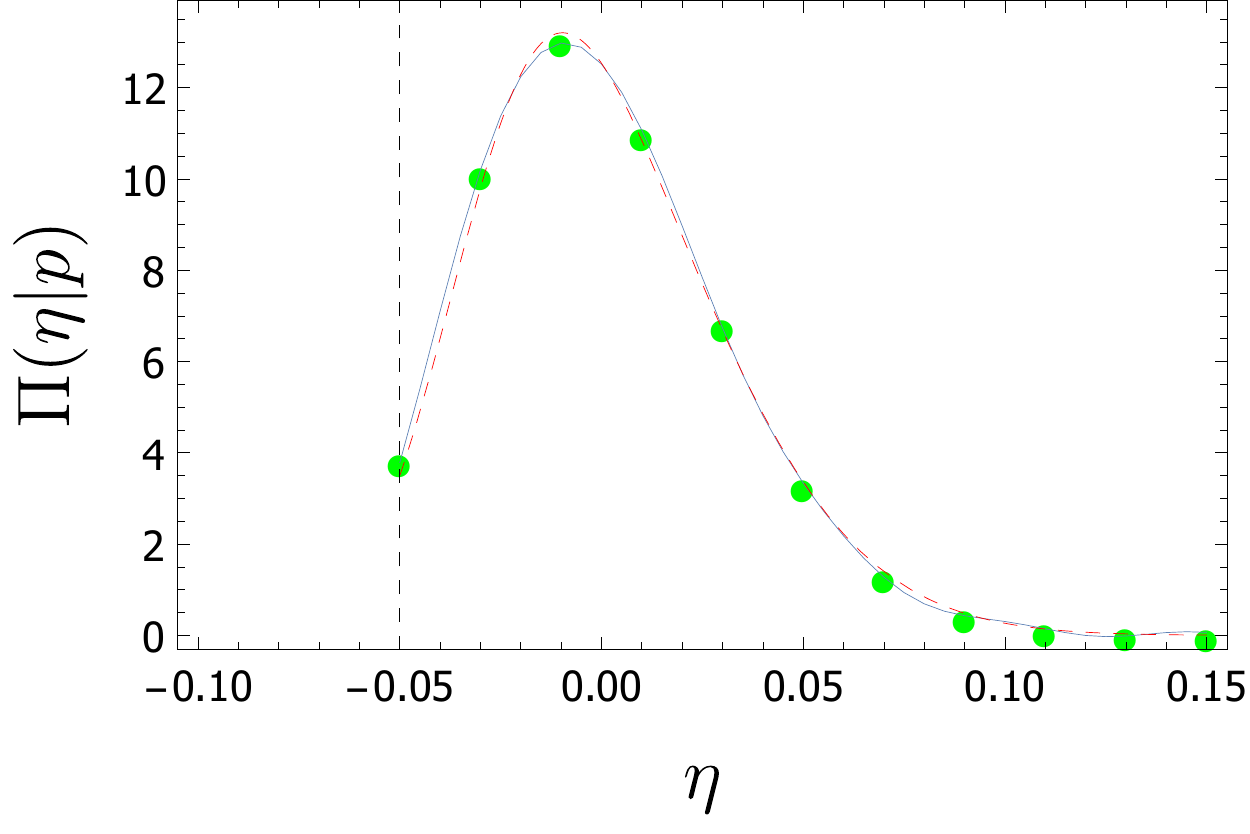}}
\caption{Blue solid line: probability distribution obtained from the first 150 moments of the mesoscopic expansion; red dashed line: skew-normal distribution; green circles: binomial distribution from the microscopic Markov chain. For the continuous distributions, the top and bottom panels correspond, respectively, to the truncated and renormalized and to the censored versions. The parameter values are $N=50$ and $p=0.05$.}
\label{fig:prob_dist_trunc}
\end{figure}

\begin{figure}[h!]
\centering
\includegraphics[scale=.63]{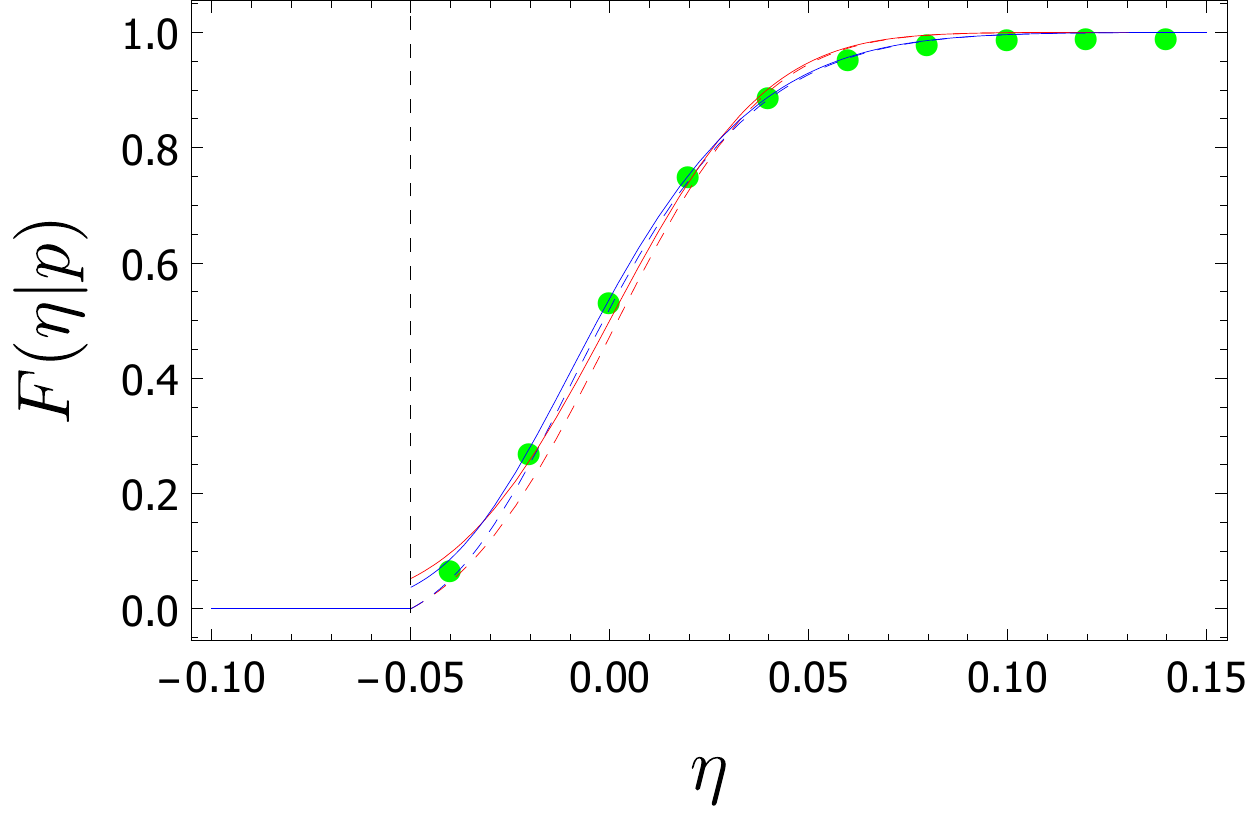}
\caption{Cumulative distribution, $F(\eta\vert p)$, for $Q_{nm}$ (green dots), and the Gaussian (red) and skew-normal (blue) parametrizations of $\Pi(\eta \vert p)$. The dashed and continuous lines correspond to the truncated and censored distributions, respectively. The parameter values are $N=50$ and $p=0.05$.}
\label{fig:cdfs}
\end{figure}

There are two ways to achieve this which are discussed in the literature~\cite{maddala_83,greene_12}. In the first, one constructs the ``truncated distribution'', which consists of truncating the noise distribution at $\eta=-p$ and $\eta=1-p$ and renormalizing it so that the area under the curve is again $1$. The second approach, consists of ``censoring'' the distribution so that $\eta$ is drawn from a distribution where the probability that it is exactly $-p$ (resp. $1-p$) corresponds to the probability that $\eta\leq -p$ (resp. $\eta \geq 1-p$) in the original distribution. In other words, the probability mass lying outside of the unit interval is not redistributed in the whole interval \emph{via} renormalization but rather placed at the boundaries themselves. In Fig.~\ref{fig:prob_dist_trunc} we compare the microscopic form of the distribution $Q_{nm}$ to the truncated and renormalized $\Pi(\eta|p)$, as well as its censored version, and find that in both cases---the distribution obtained with higher moments and its skew-normal parametrization---the censored distribution yields better agreement with the microscopic distribution $Q_{nm}$, an effect which is more pronounced for the skew-normal parametrization. We can further illustrate this by comparing the cumulative distribution functions of the truncated and censored distributions, which are shown in Fig.~\ref{fig:cdfs} both for the Gaussian and skew-normal distributions.


We end this section with a summary of the methods we have suggested to treat the problem of trajectories leaving the interval on which the model is defined. The method of \textit{random replacement} employs the stochastic difference equation used previously \cite{challenger_13,challenger_14,parra_14}, which is summarized in Eqs.~(\ref{stoch_diff_eqn}) and (\ref{correlator}). When a trajectory leaves the unit interval it is returned to the interval at a randomly chosen position. The \textit{truncated Gaussian} method uses the same Gaussian distribution for the noise, truncated and renormalized so that trajectories cannot leave the interval. The \textit{censored Gaussian} method again uses the Gaussian noise, but this time the total probability to leave the interval through the lower (or upper) boundary is instead placed at that boundary. The methods of censoring and truncating the distributions are also used with the skew-normal distribution introduced in Sec~\ref{sec:meso} for the methods of \textit{censored skew-normal} and \textit{truncated skew normal}. This distribution is parameterized as shown in Eq.~(\ref{para_skew_dist}). We now compare these different strategies to each other and to the results obtained from the original, microscopic, model.

\section{Results for the quasistationary probability distribution}
\label{sec:results}

In the previous section we have described a number of different descriptions of the noise in the stochastic difference equation. These have involved placing restrictions on the noise, so that trajectories remain the unit interval. Allied to this, we have examined the non-Gaussian nature of the noise near the boundary, and have proposed including higher-order jump moments, parameterized through the use of a skew-normal distribution to describe the noise. For a range of $(N,\lambda)$ we will now show results from all five prescriptions for the noise, each designed to prevent trajectories leaving the interval. 

\begin{figure*}
\centering
\subfigure{\includegraphics[scale=.47]{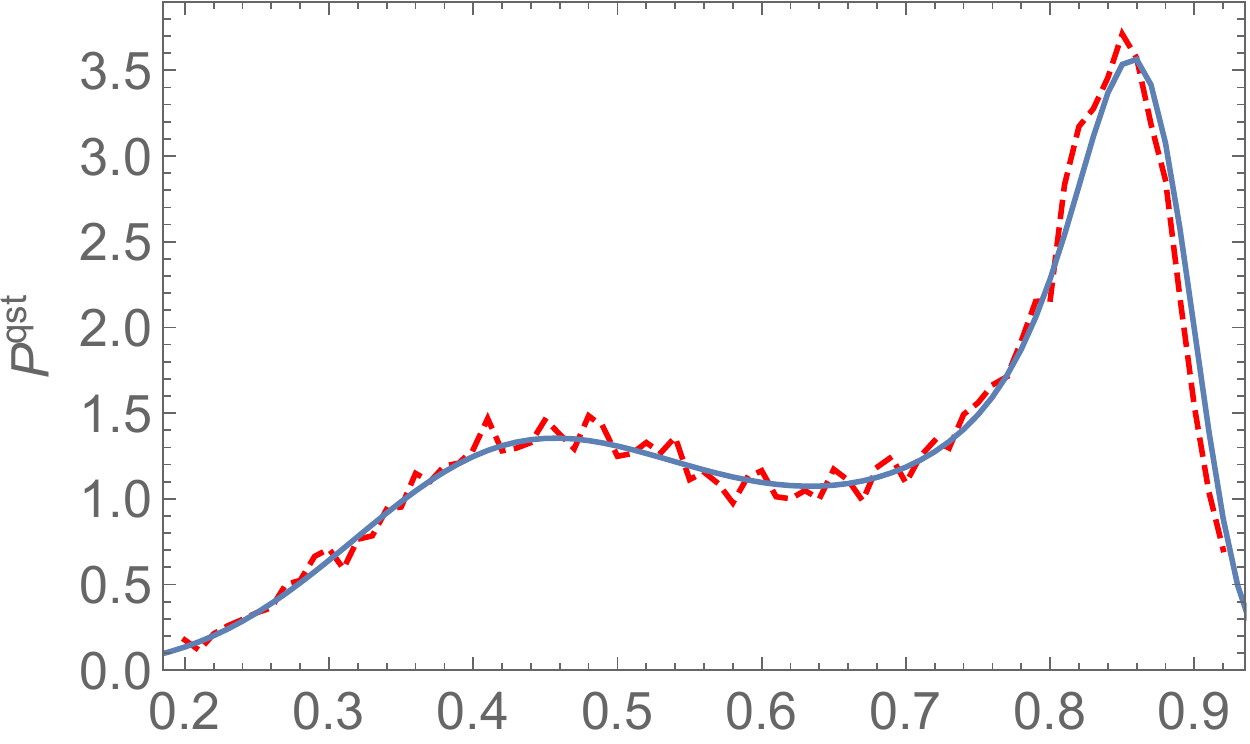}}
\subfigure{\includegraphics[scale=.44]{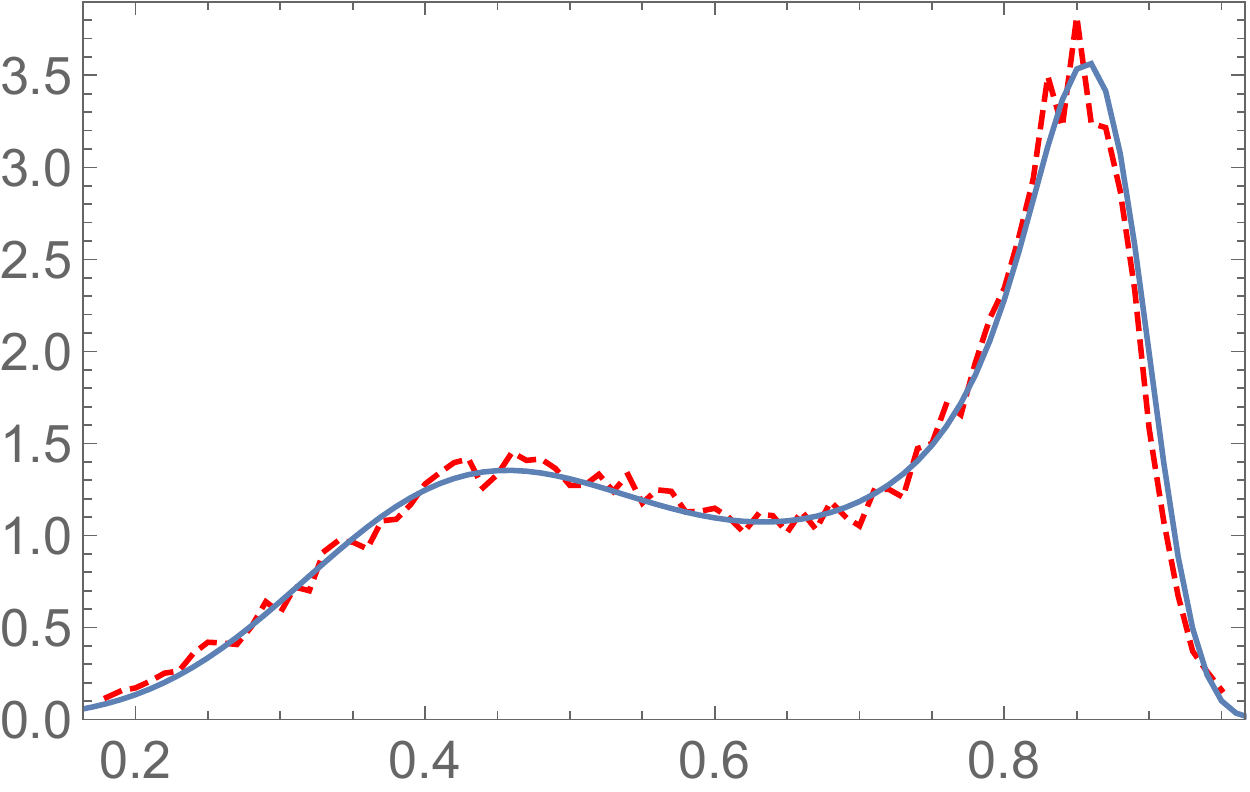}}
\subfigure{\includegraphics[scale=.44]{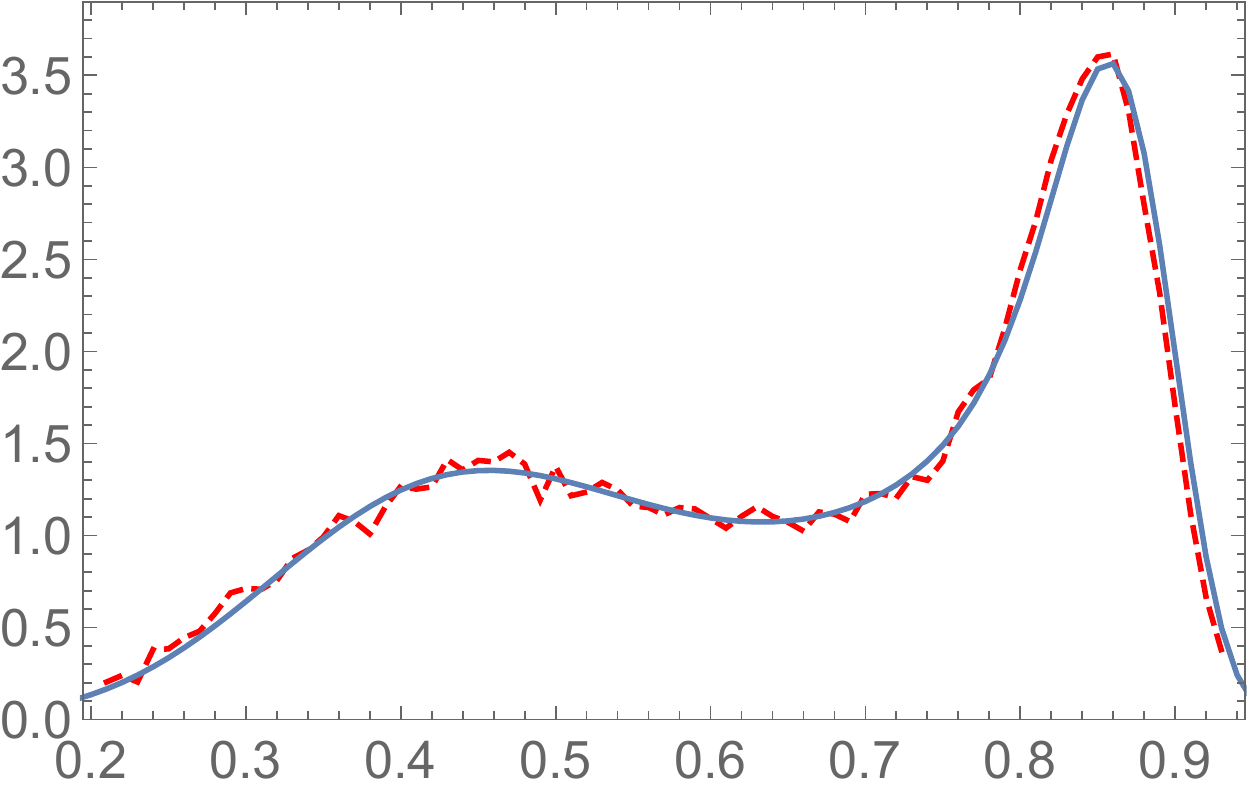}}\\
\subfigure{\includegraphics[scale=.47]{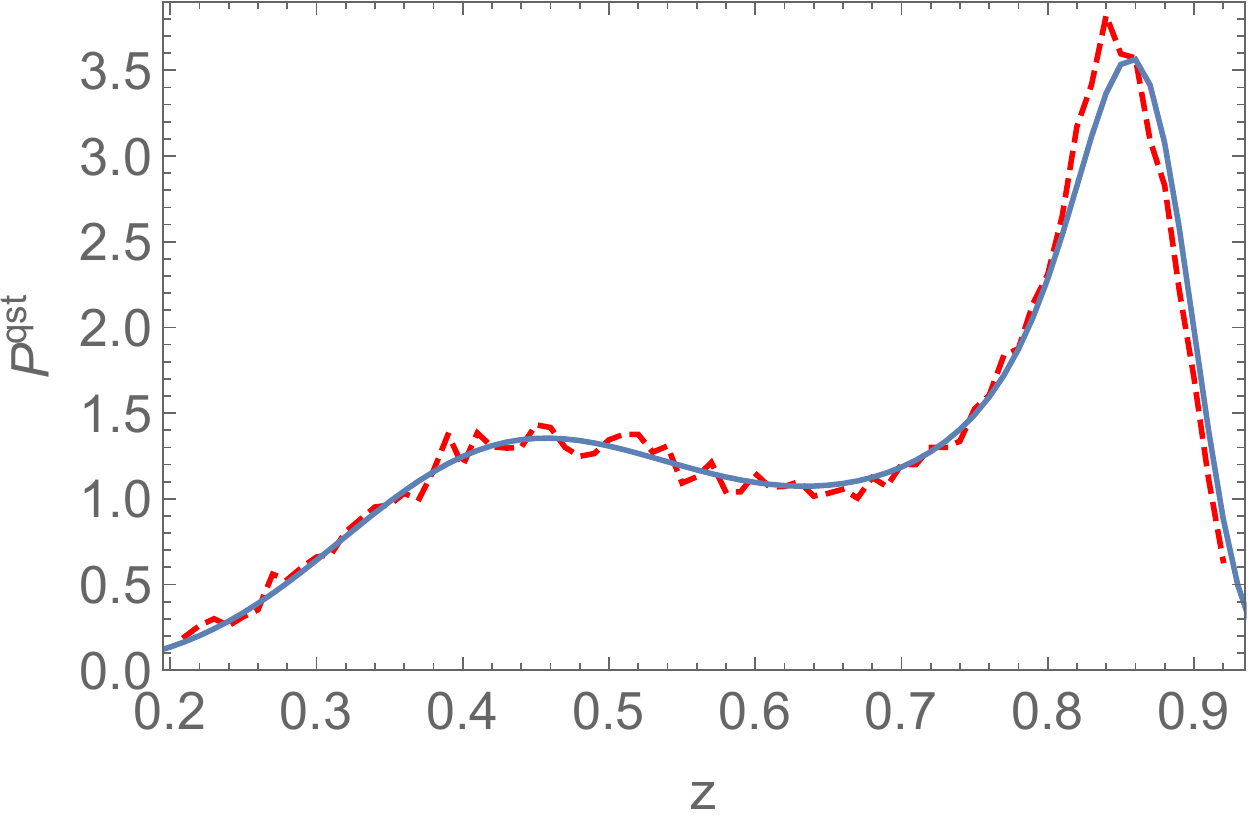}}
\subfigure{\includegraphics[scale=.44]{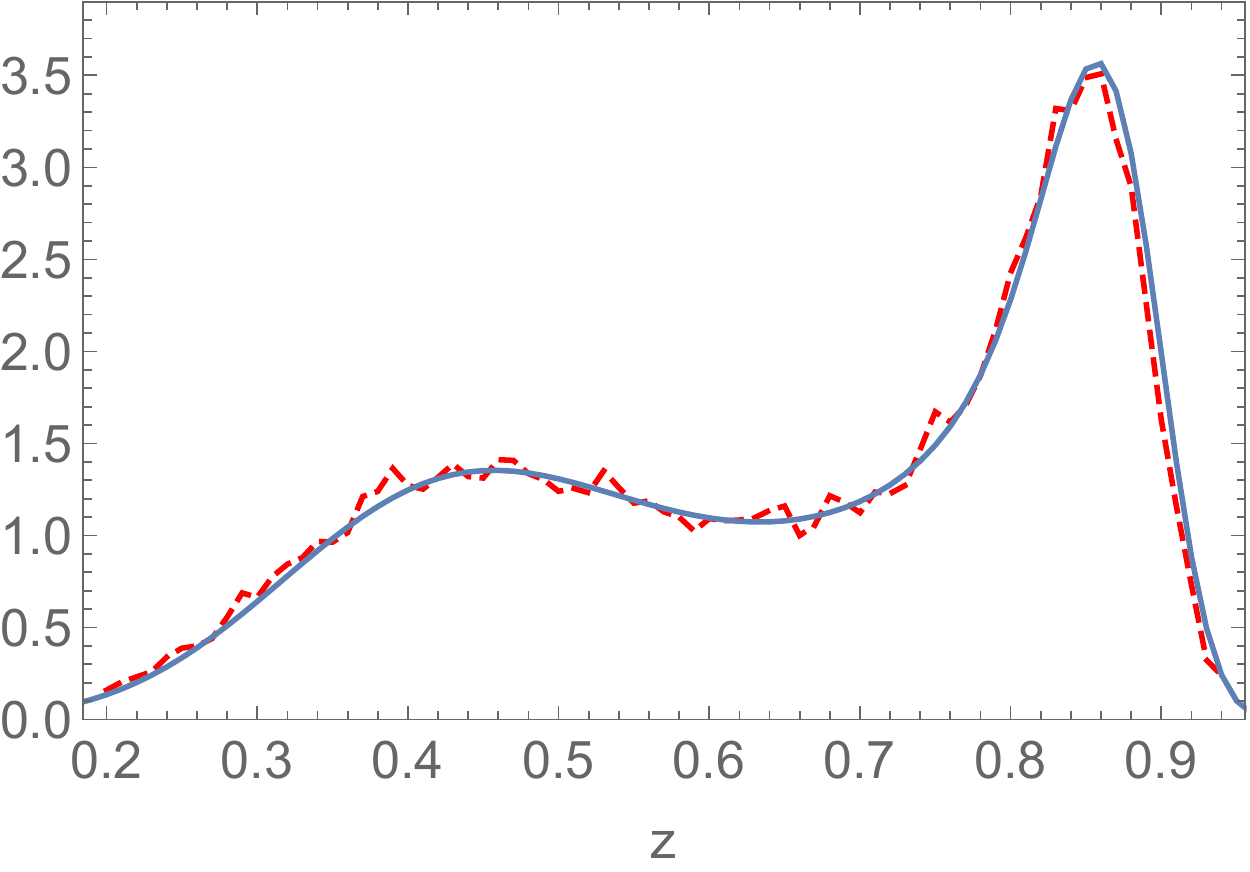}}
\subfigure{\includegraphics[scale=.44]{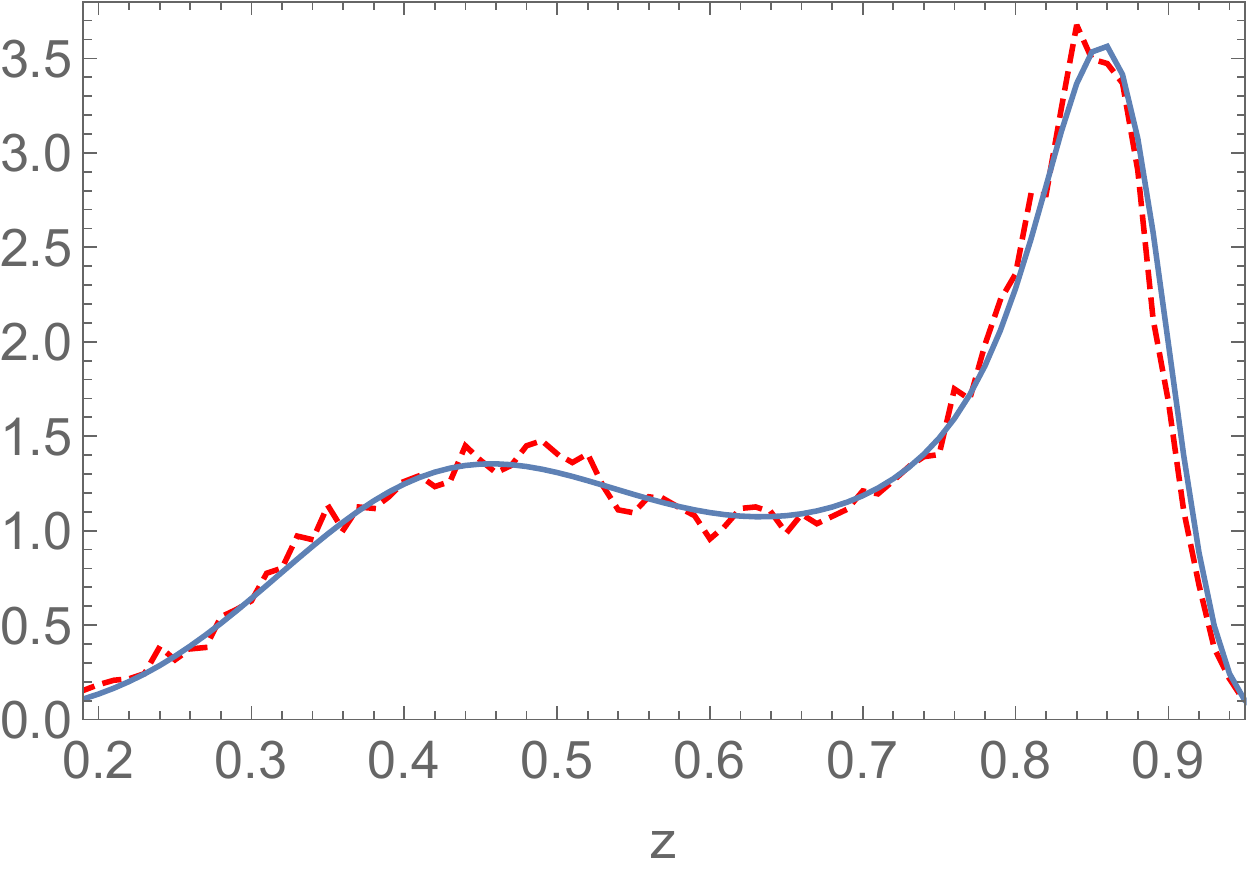}}
\caption{Quasistationary distribution of the Markov chain, obtained as the eigenvector corresponding to the largest eigenvalue of $\hat{\bm {Q}}$ (blue solid line), compared to the same result from the SDE (red dashed line) using different methods of simulation. Top row (from left to right): random replacement, truncated Gaussian, and truncated skew-normal. Bottom row (from left to right): censored Gaussian, censored skew-normal, and the modified censored Gaussian. 
Parameter values: $\lambda=3.5$, $N=100$.}
\label{fig:qsd_meso_1}
\end{figure*}

\begin{figure*}
\centering
\subfigure{\includegraphics[scale=.48]{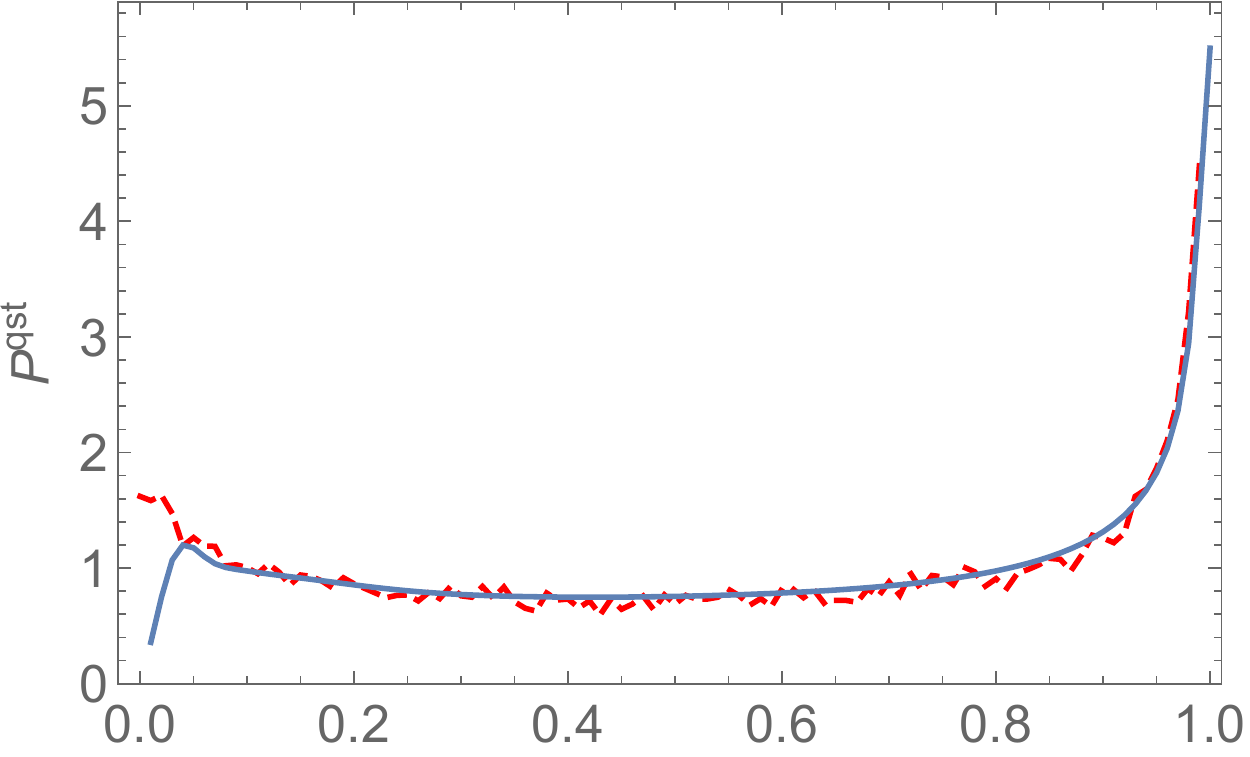}}
\subfigure{\includegraphics[scale=.45]{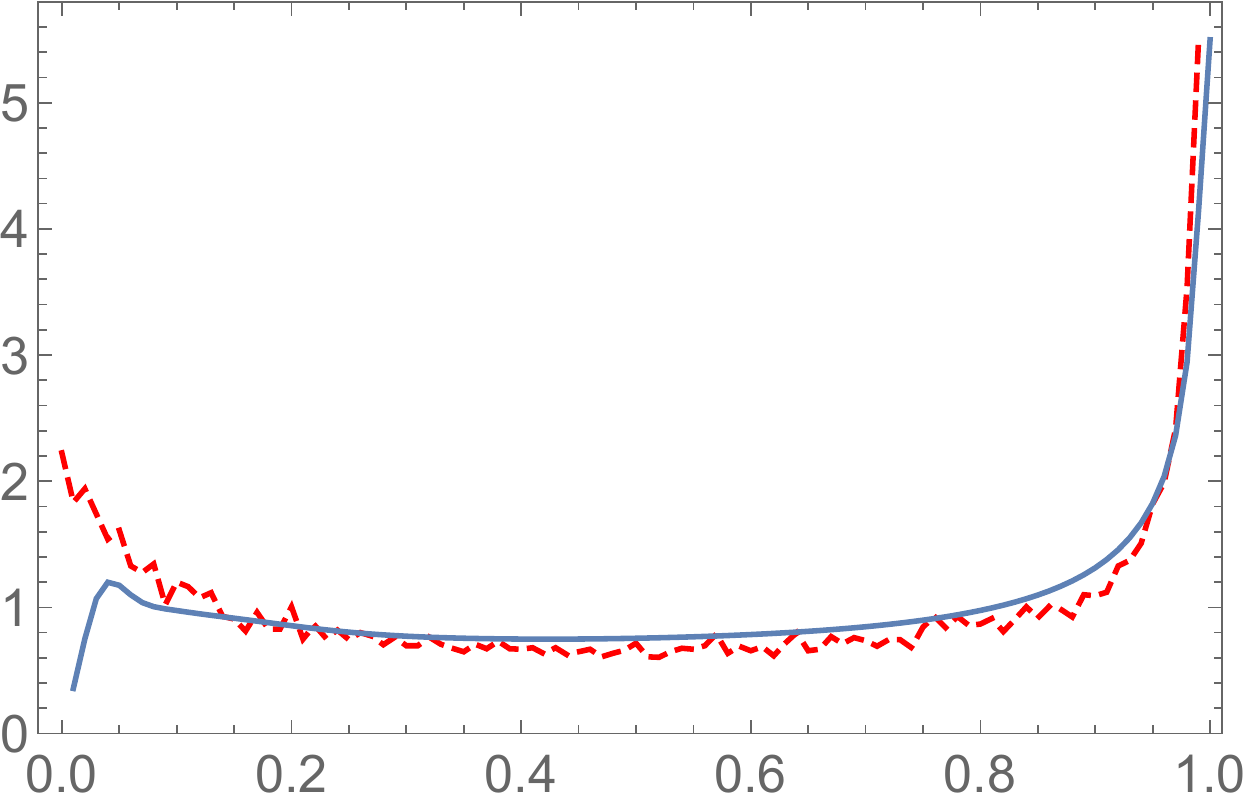}}
\subfigure{\includegraphics[scale=.45]{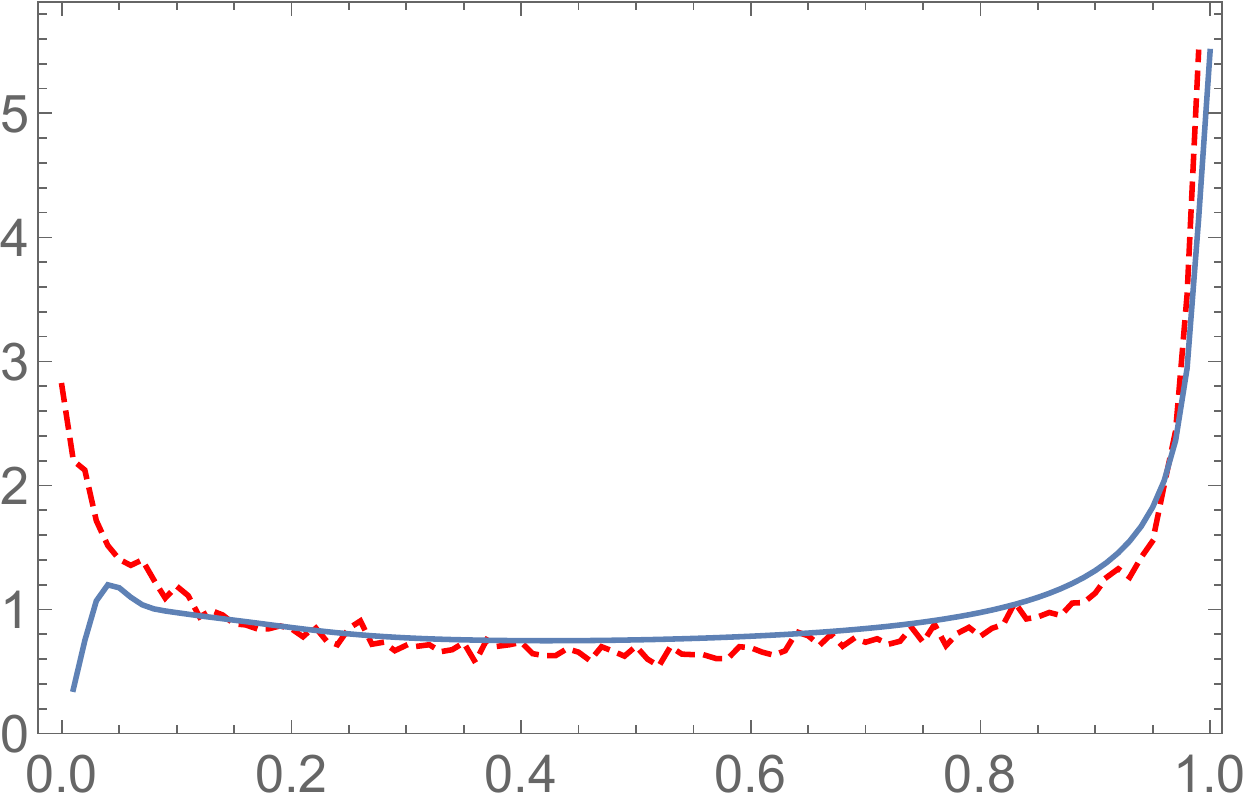}}\\
\subfigure{\includegraphics[scale=.48]{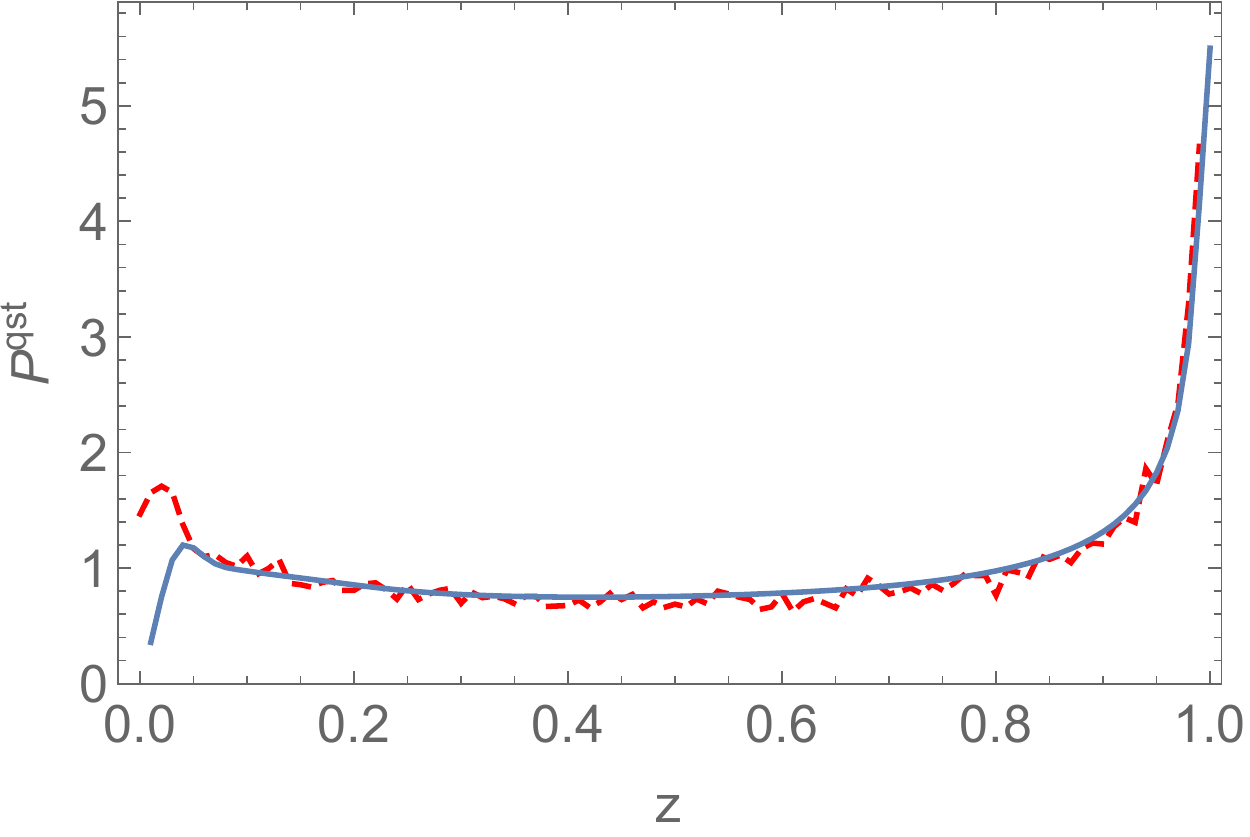}}
\subfigure{\includegraphics[scale=.45]{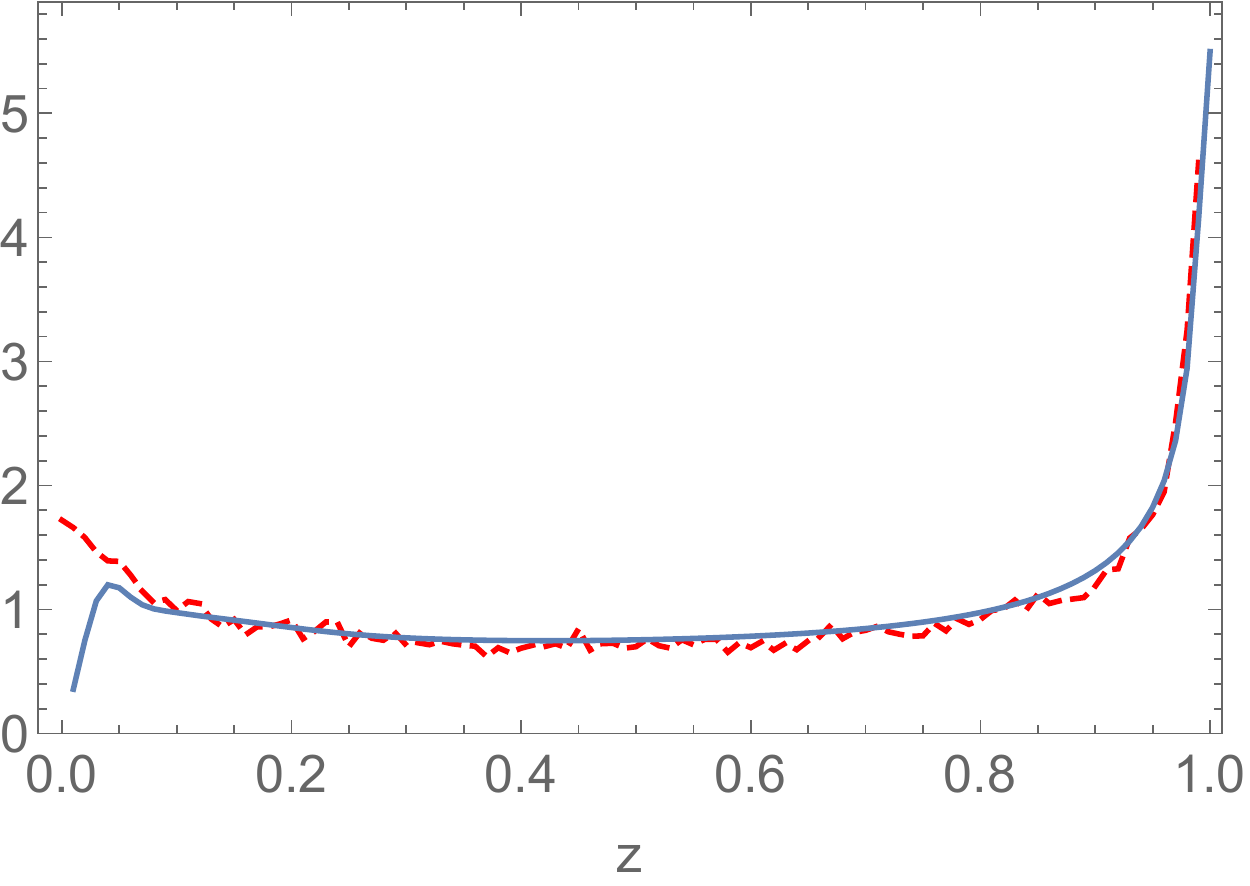}}
\subfigure{\includegraphics[scale=.45]{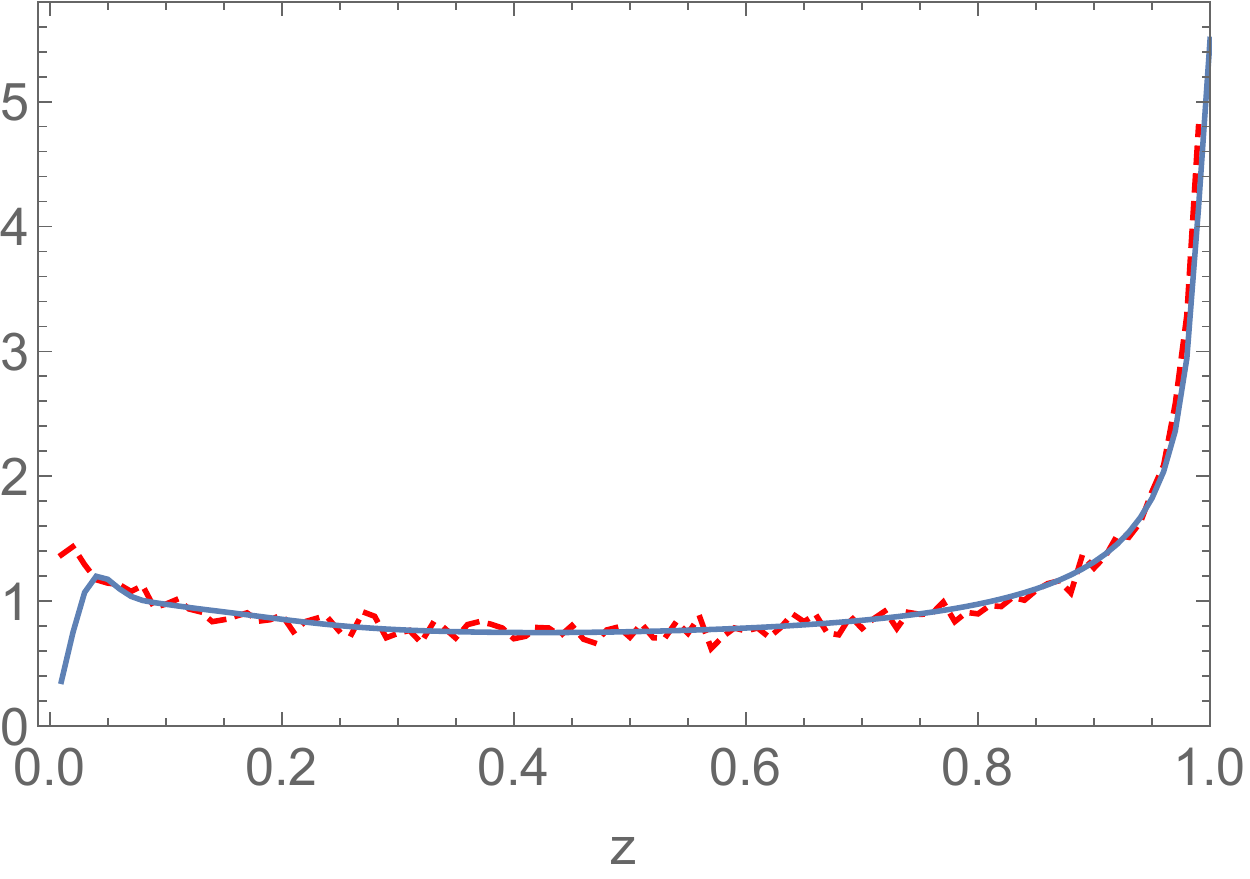}}
\caption{Quasistationary distribution of the Markov chain, obtained as the eigenvector corresponding to the largest eigenvalue of $\hat{\bm {Q}}$ (blue solid line), compared to the same result from the SDE (red dashed line) using different methods of simulation. Top row (from left to right): random replacement, truncated Gaussian, and truncated skew-normal. Bottom row (from left to right): censored Gaussian, censored skew-normal, and the modified censored Gaussian.
Parameter values: $\lambda=3.99$, $N=100$.}
\label{fig:qsd_meso_2}
\end{figure*}

\begin{figure*}
\centering
\subfigure{\includegraphics[scale=.48]{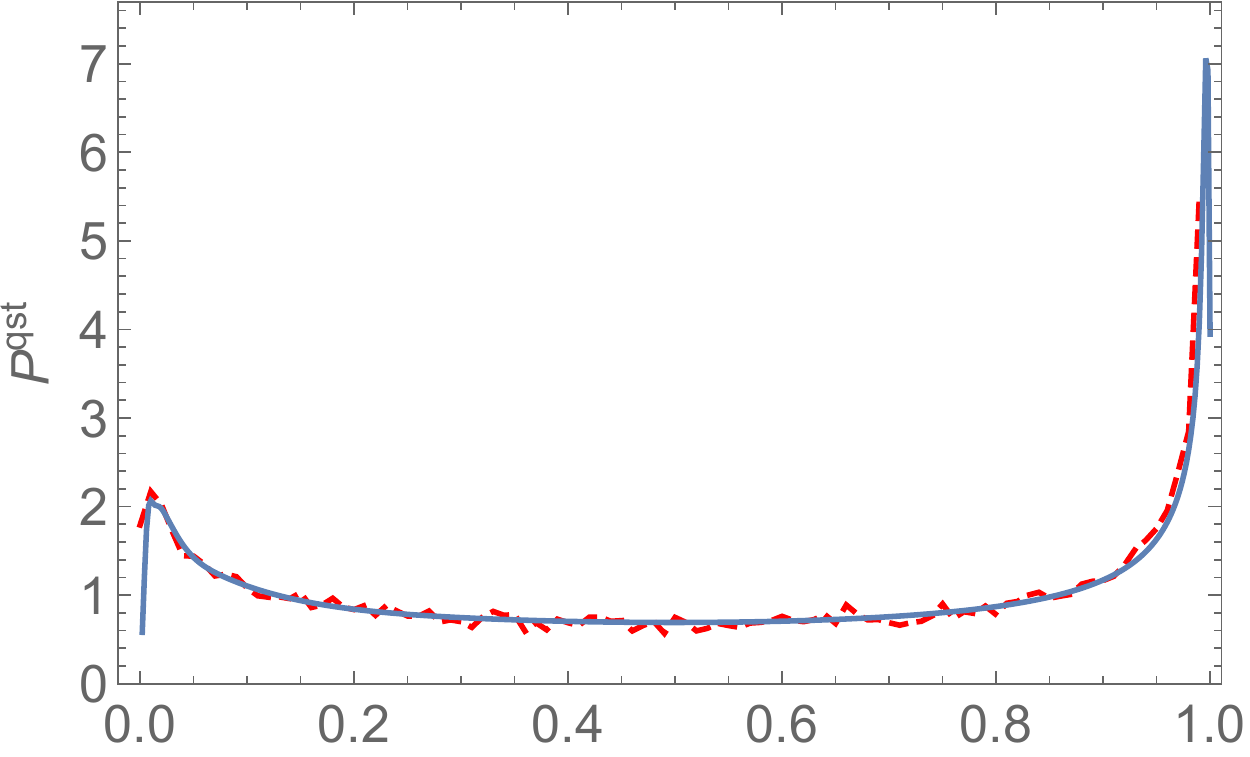}}
\subfigure{\includegraphics[scale=.45]{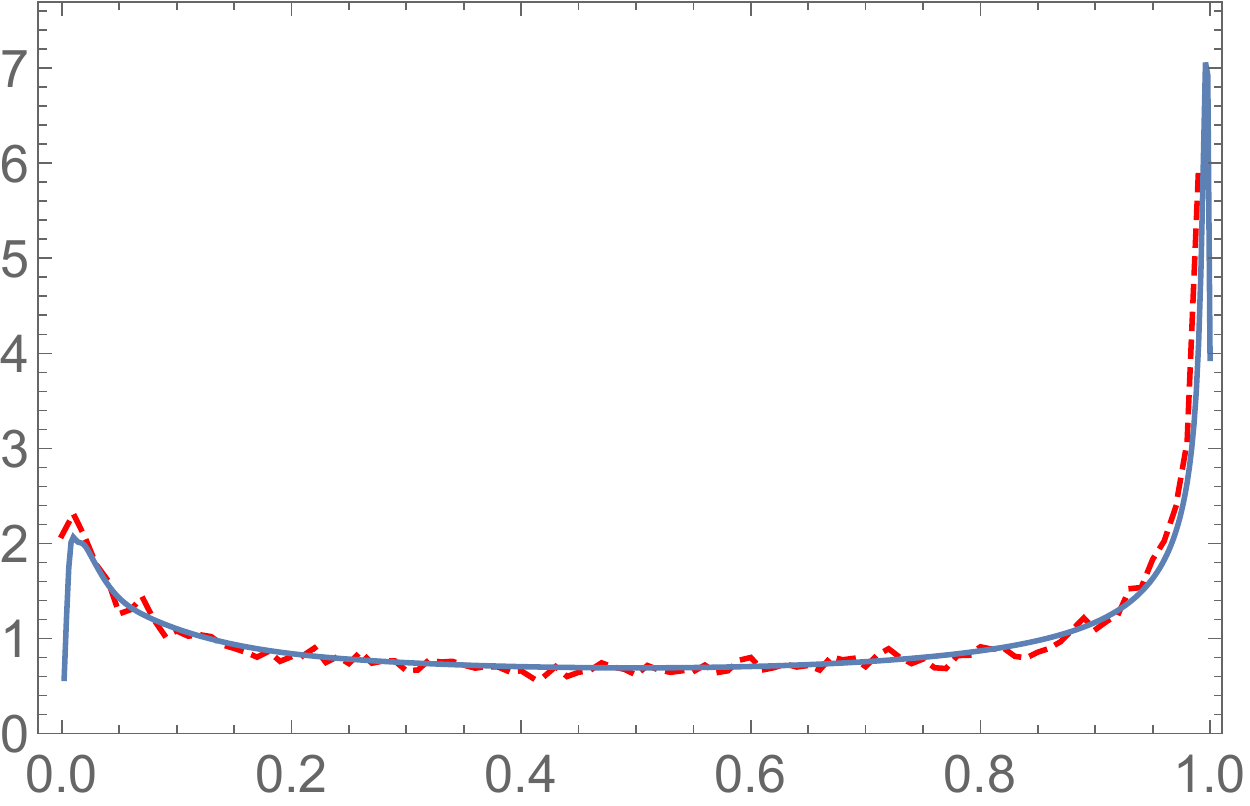}}
\subfigure{\includegraphics[scale=.45]{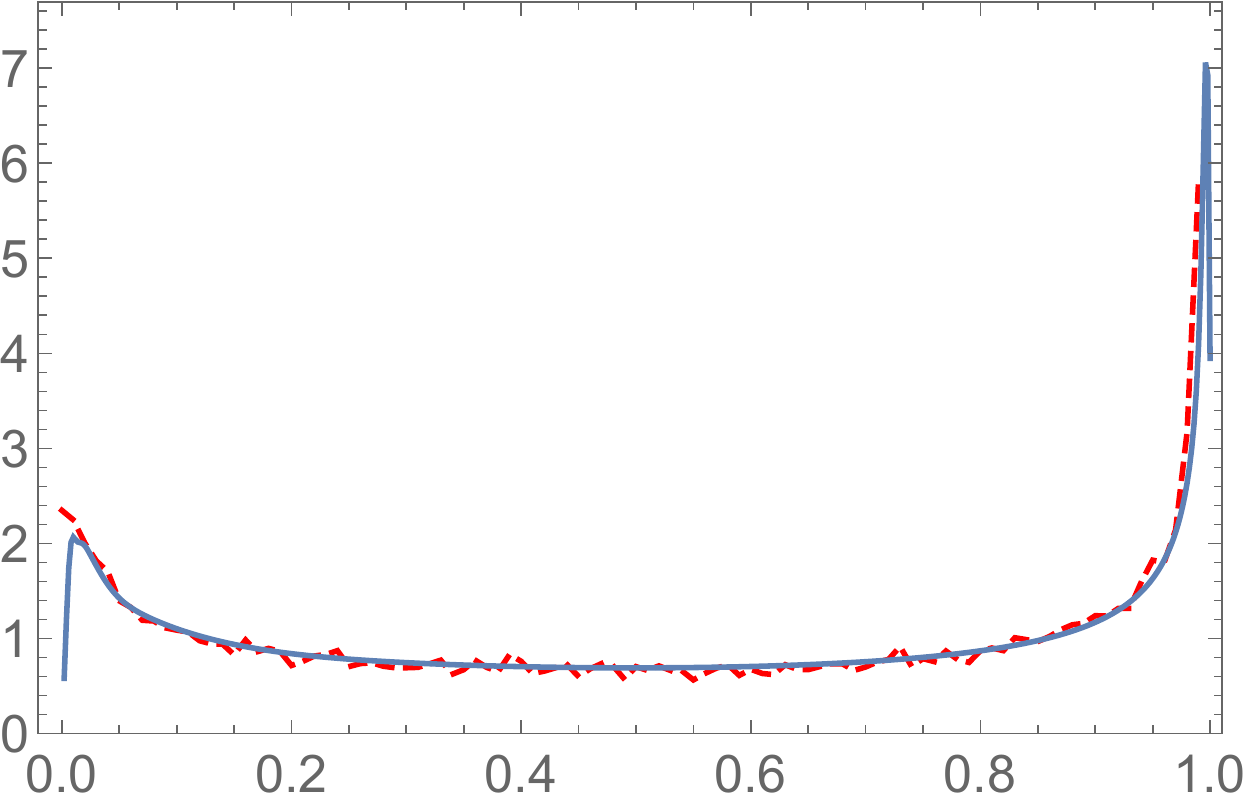}}\\
\subfigure{\includegraphics[scale=.48]{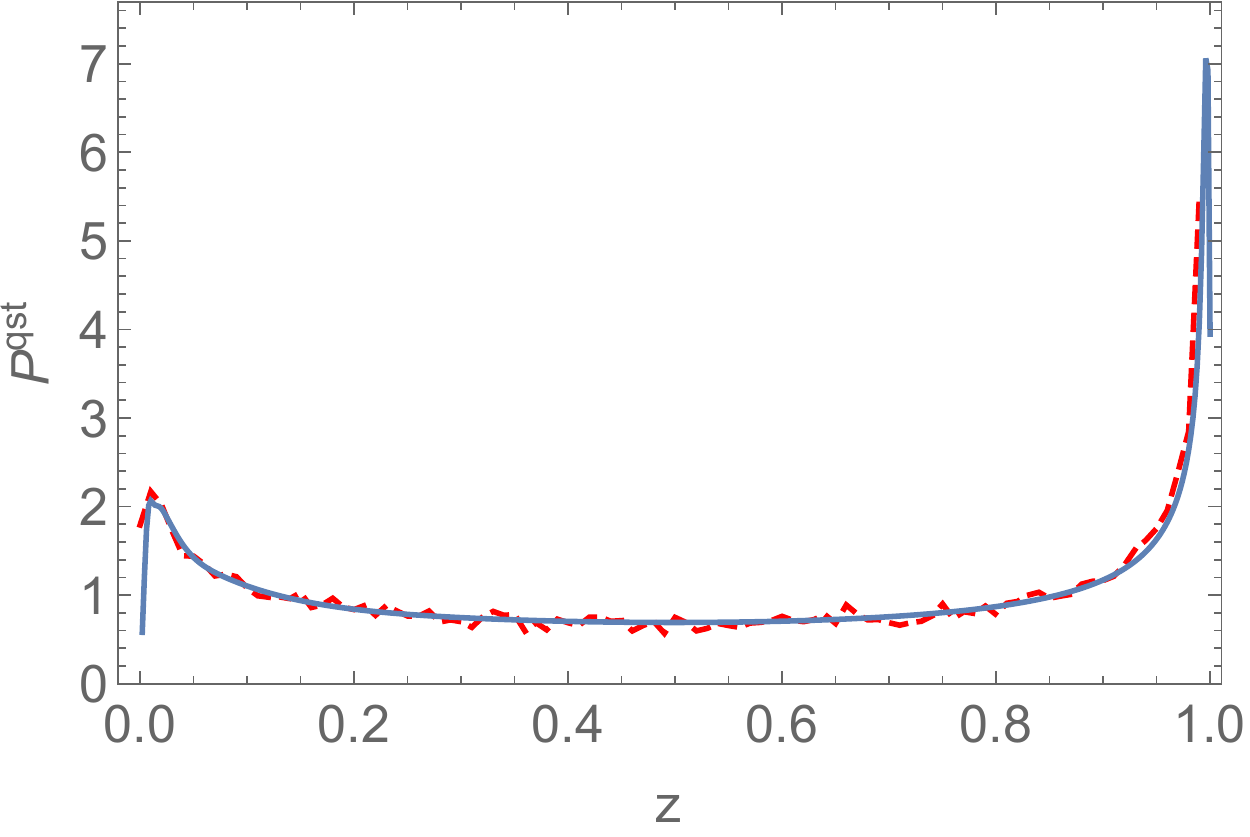}}
\subfigure{\includegraphics[scale=.45]{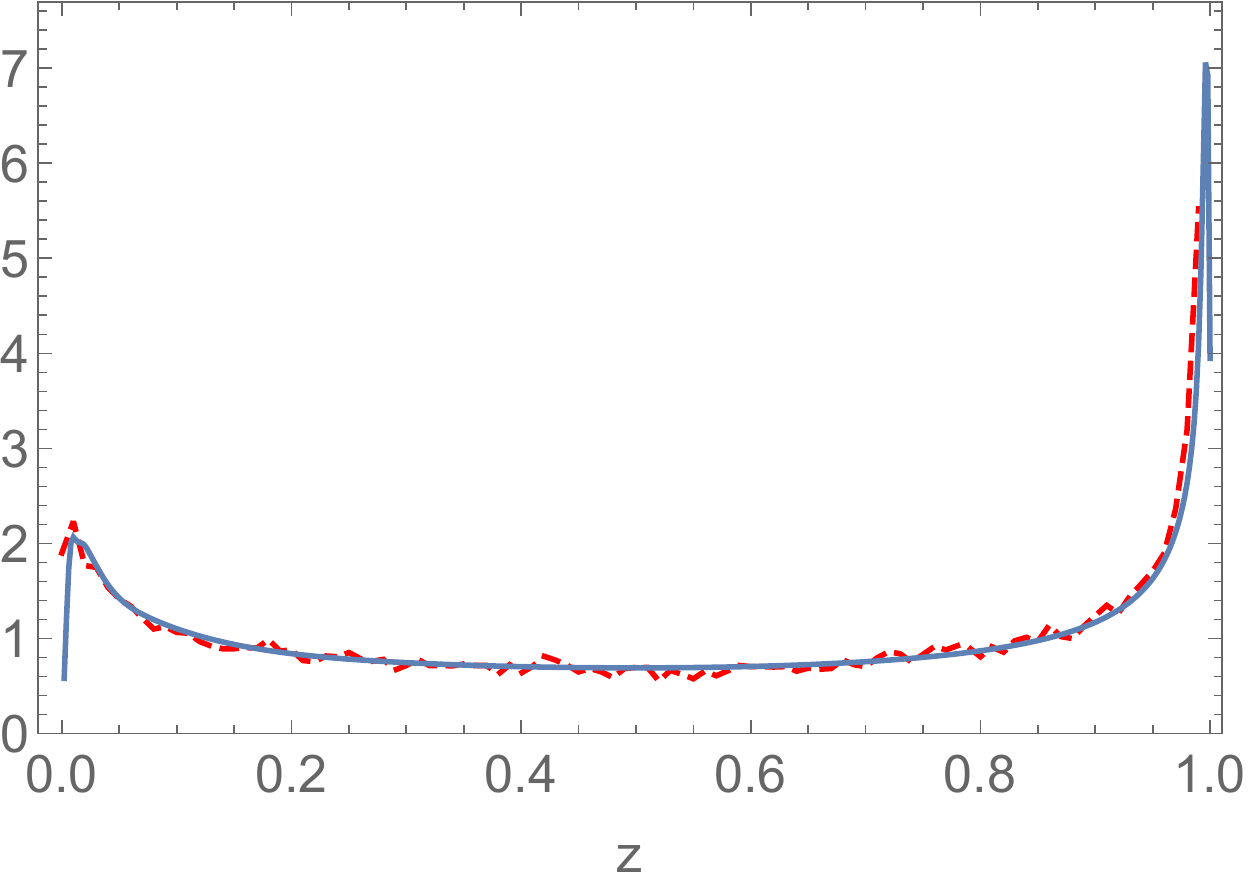}}
\subfigure{\includegraphics[scale=.45]{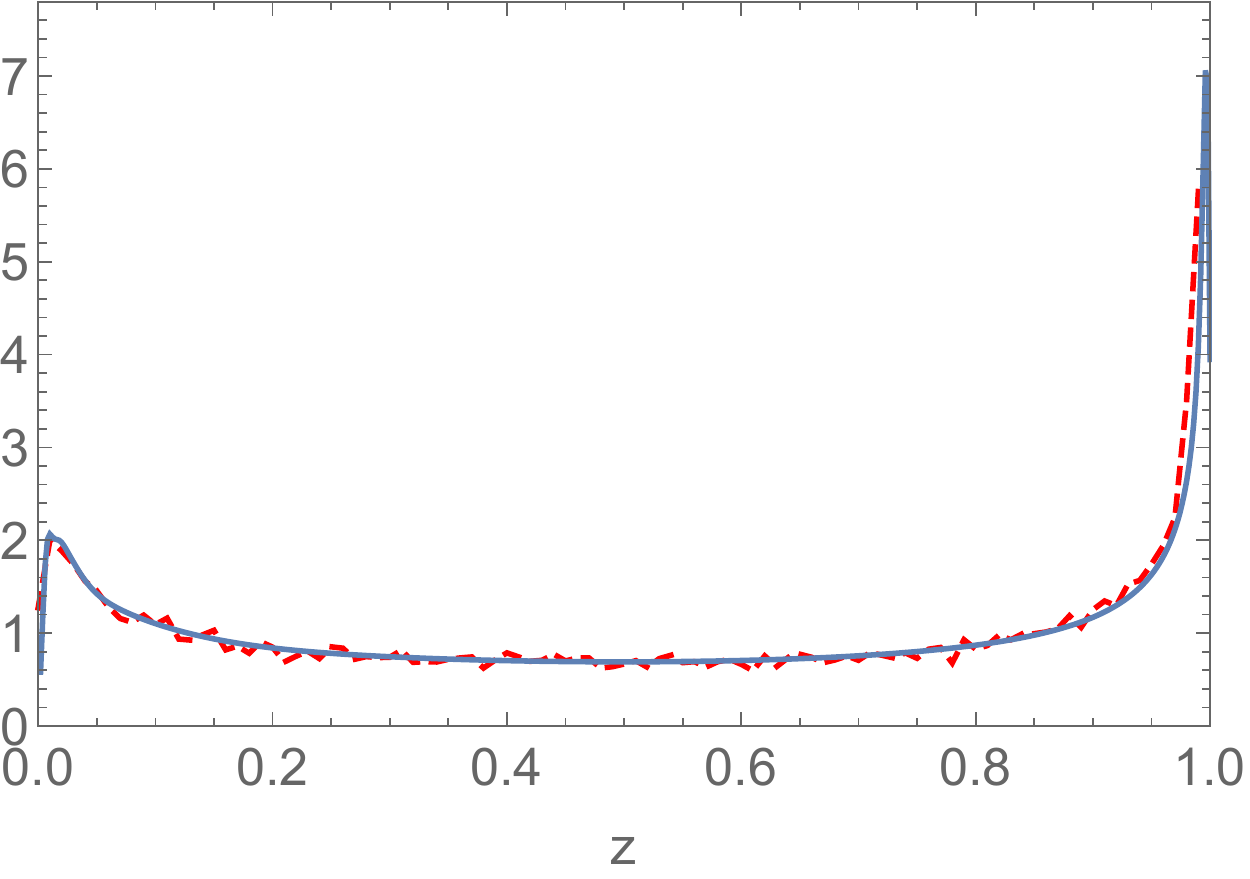}}
\caption{Quasistationary distribution of the Markov chain, obtained as the eigenvector corresponding to the largest eigenvalue of $\hat{\bm {Q}}$ (blue solid line), compared to the same result from the SDE (red dashed line) using different methods of simulation. Top row (from left to right): random replacement, truncated Gaussian, and truncated skew-normal. Bottom row (from left to right): censored Gaussian, censored skew-normal, and the modified censored Gaussian.
Parameter values: $\lambda=3.99$, $N=500$.}
\label{fig:qsd_meso_3}
\end{figure*}

\begin{figure*}
\centering
\subfigure{\includegraphics[scale=.48]{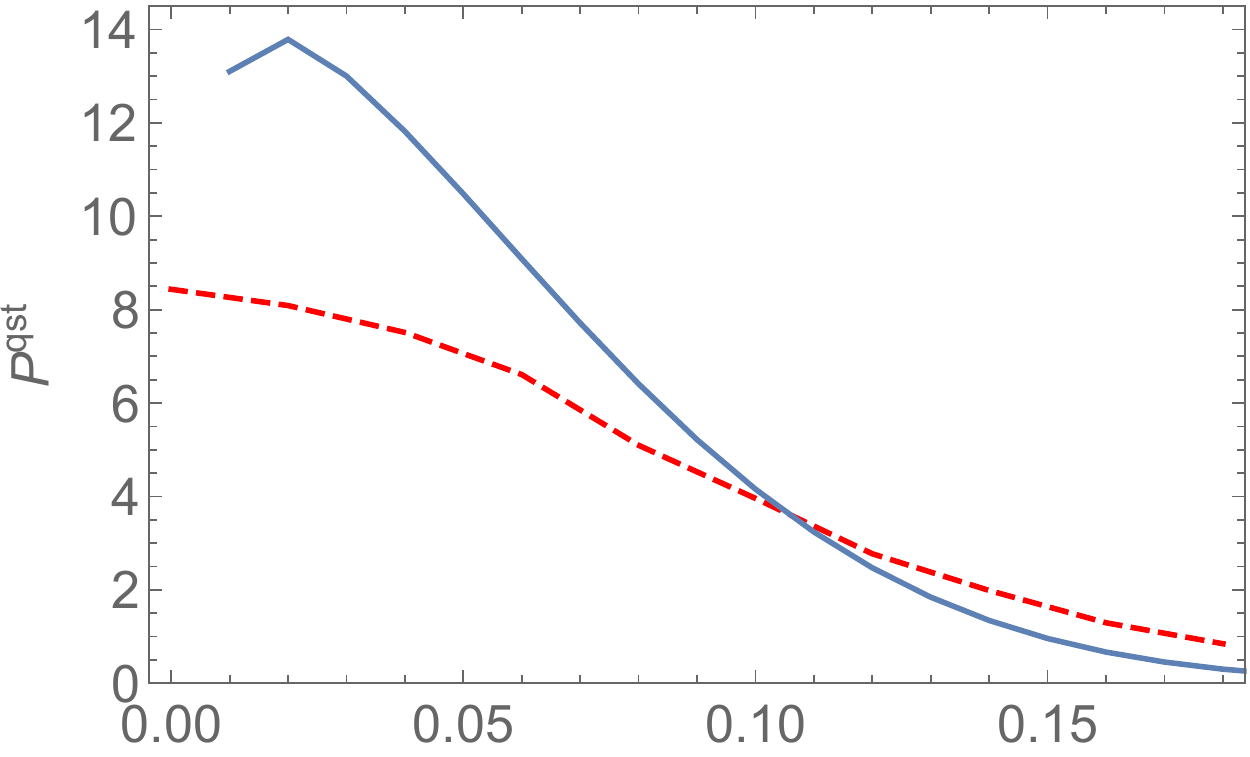}}
\subfigure{\includegraphics[scale=.45]{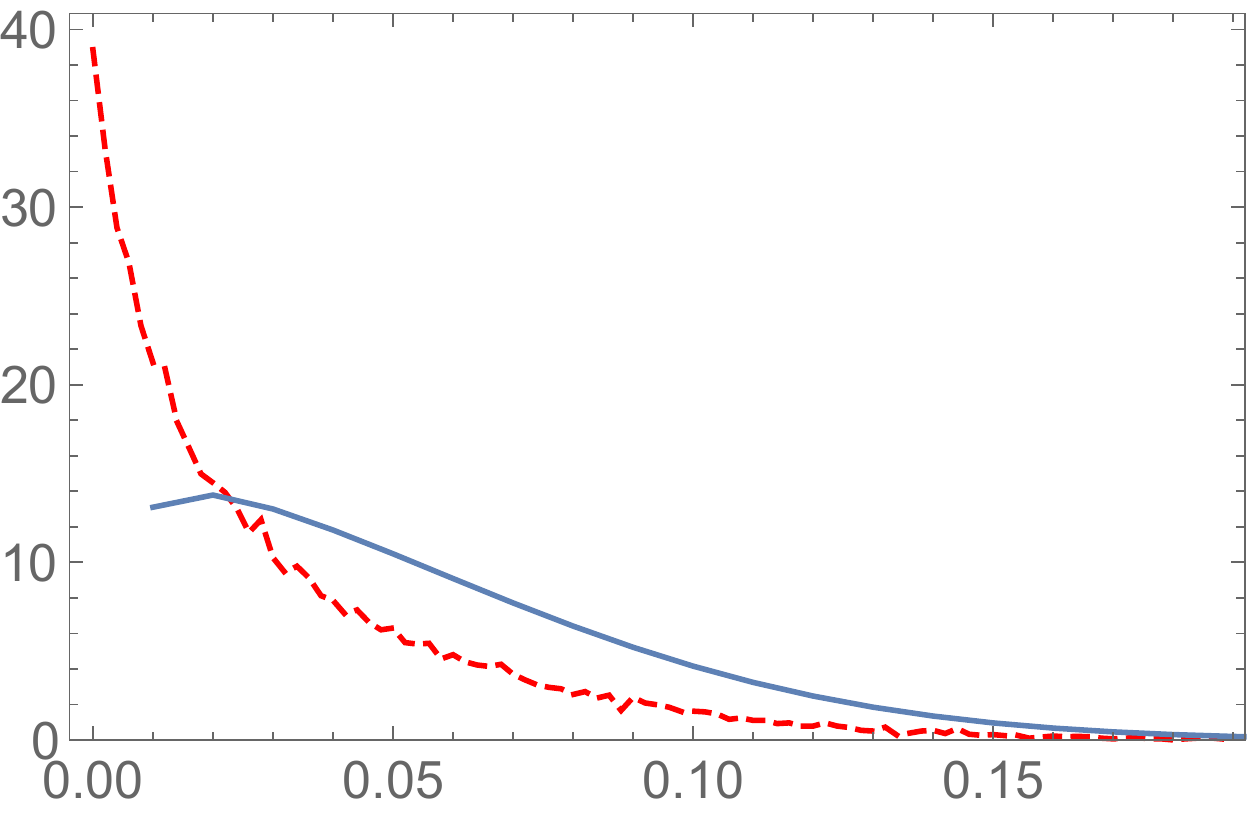}}
\subfigure{\includegraphics[scale=.45]{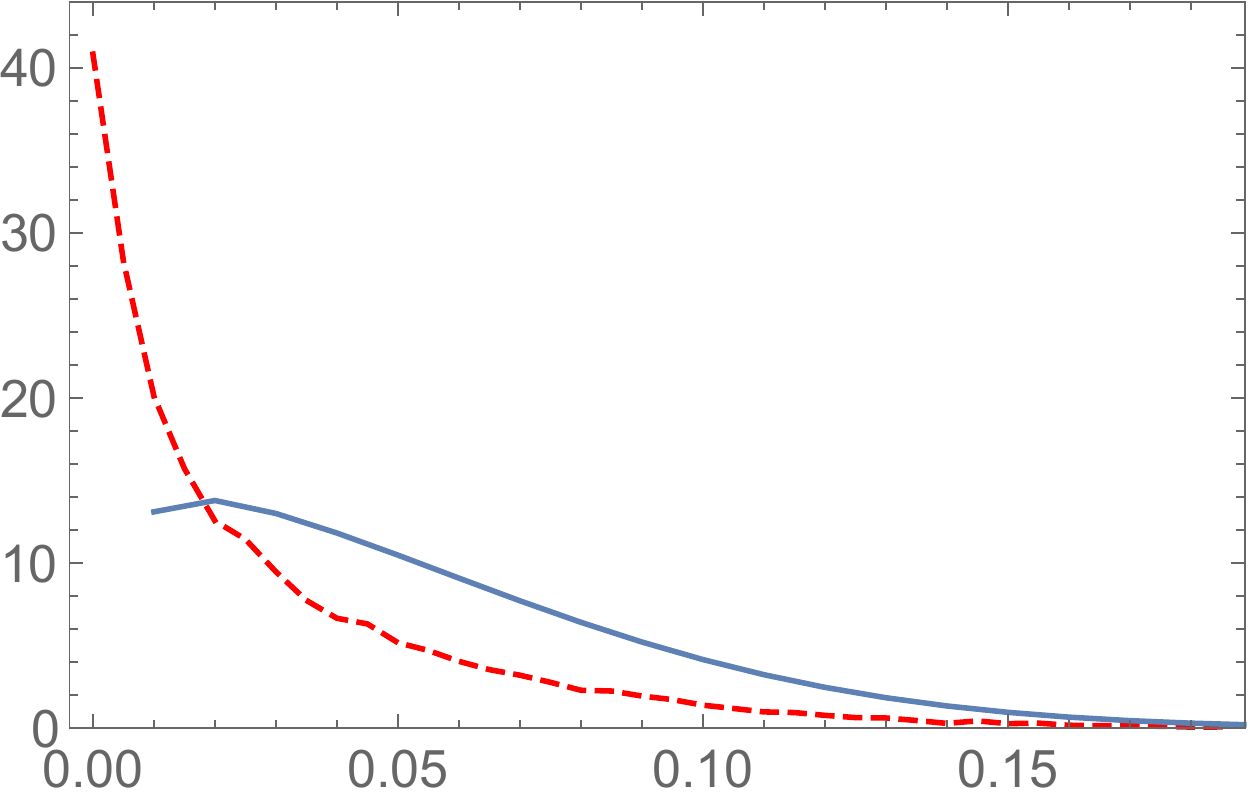}}\\
\subfigure{\includegraphics[scale=.48]{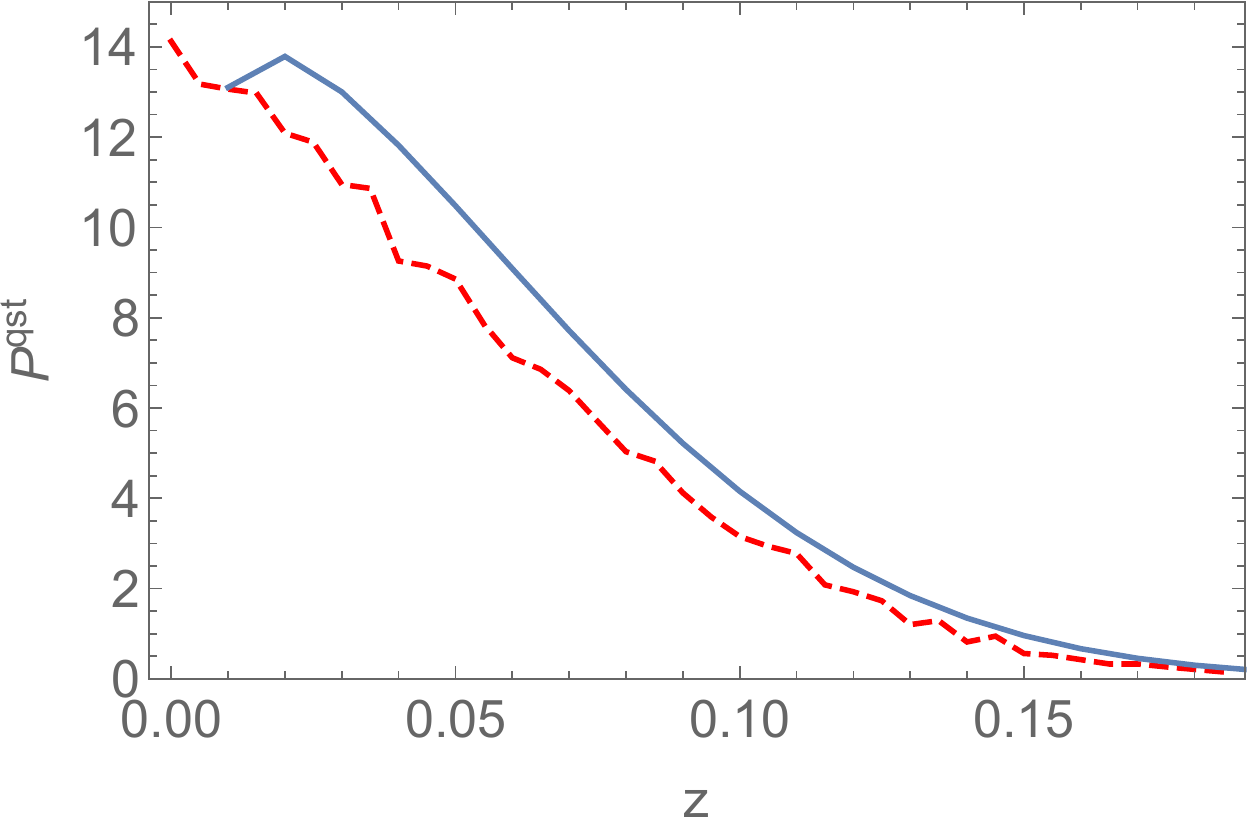}}
\subfigure{\includegraphics[scale=.45]{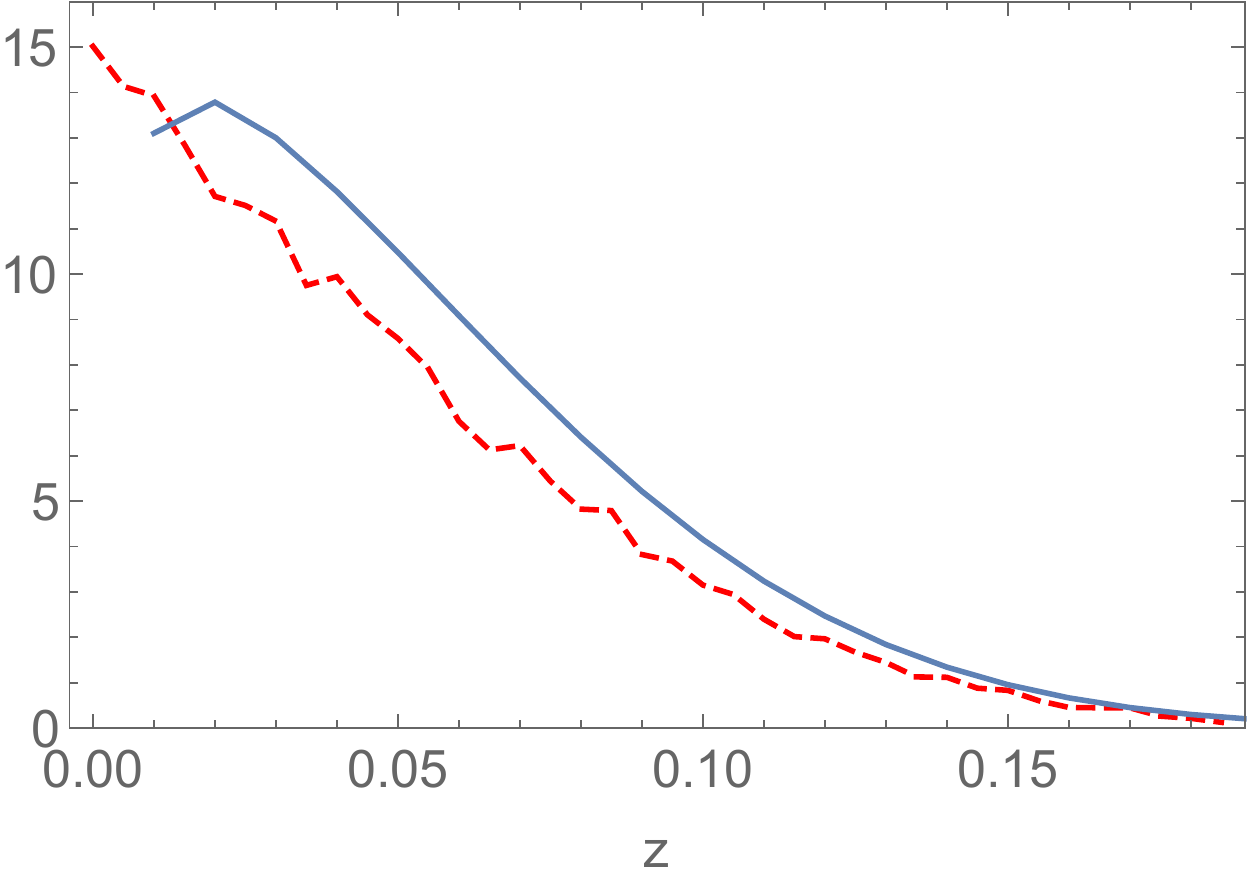}}
\subfigure{\includegraphics[scale=.45]{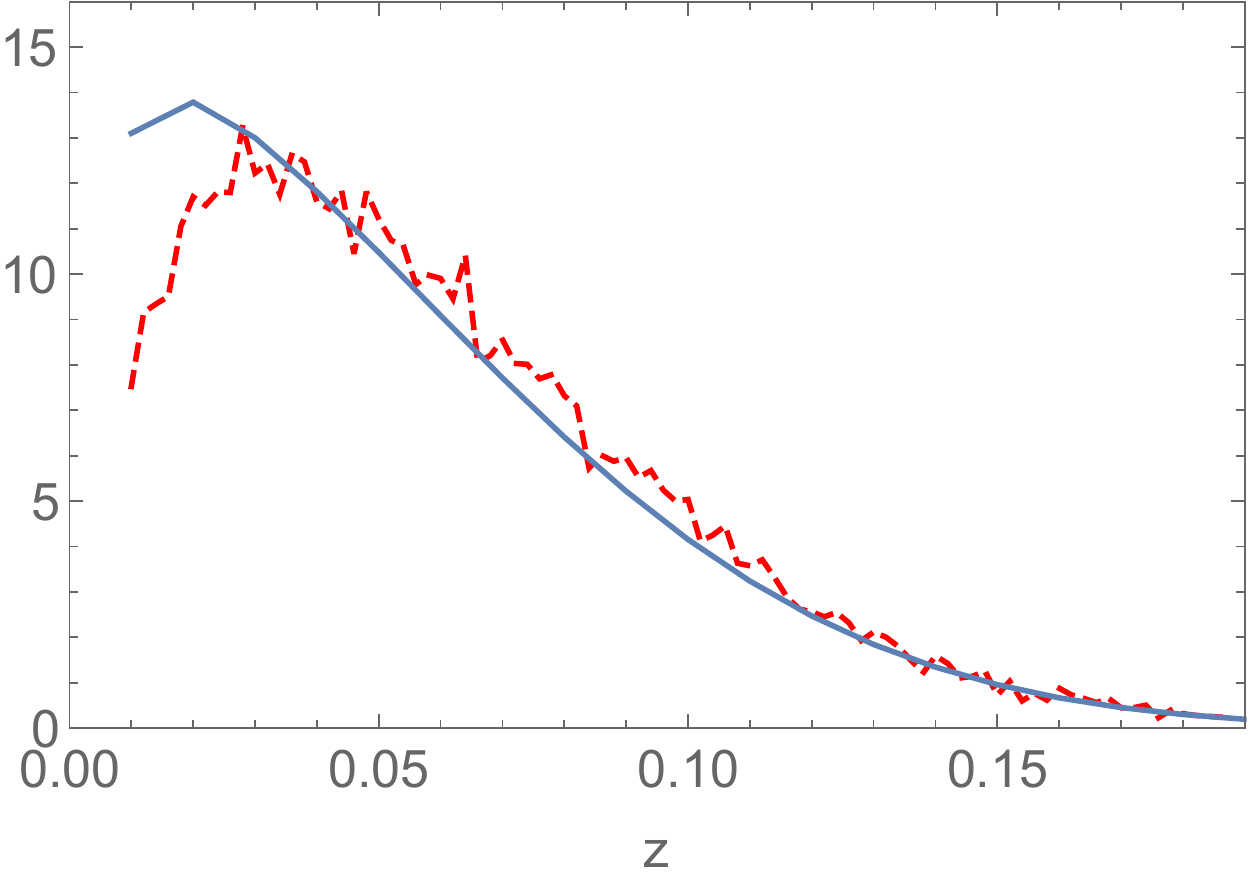}}
\caption{Quasistationary distribution of the Markov chain, obtained as the eigenvector corresponding to the largest eigenvalue of $\hat{\bm {Q}}$ (blue solid line), compared to the same result from the SDE (red dashed line) using different methods of simulation. Top row (from left to right): random replacement, truncated Gaussian, and truncated skew-normal. Bottom row (from left to right): censored Gaussian, censored skew-normal, and the modified censored Gaussian.
Parameter values: $\lambda=1.01$, $N=100$.}
\label{fig:qsd_meso_4}
\end{figure*}

In Figs.~\ref{fig:qsd_meso_1}--\ref{fig:qsd_meso_4} we compare the results for the quasistationary probability distribution for all the methods above against the desired result obtained from the Markov chain, for the logistic map with different choices of parameters. In the case where the noise distribution is truncated, the simulations are straightforward to perform, as trajectories will not leave the unit interval. For the censored distributions, however, some trajectories will be discarded, to reflect probability building up at the boundaries. A large number of trajectories are generated, until 25000 of them survive (i.e., remain in the unit interval) until iteration number $N$. We use the system size here, since we expect the survival times to depend on the noise strength. From each of these trajectories, we take the last iterate and build a histogram as a numerical approximation of the desired probability distribution.

We find that when the dynamics stays sufficiently far from the boundaries, the simulation method employed is less important, and all of them yield a quasistationary distribution which is in good agreement with the Markov chain. An example of this is shown in Fig.~\ref{fig:qsd_meso_1}. In cases such as this one, the unbounded noise used in Eq.~(\ref{stoch_diff_eqn}) will only cause trajectories to leave the interval very rarely. Furthermore, as trajectories seldom venture close to the boundaries, the Gaussian noise in Eq.~(\ref{stoch_diff_eqn}) is a very good approximation of the microscopic dynamics. As the map spends more and more time near the boundaries, however, we observe a clearer difference between the various methods. In Figs.~\ref{fig:qsd_meso_2} and \ref{fig:qsd_meso_3} we examine the case where almost the entire interval is being explored, for two choices of $N$. For the smaller value of $N$, shown in Fig.~\ref{fig:qsd_meso_2}, all methods overestimate the amount of time spent near the boundary at zero, although the random replacement method and the censored distributions perform better than the truncated distributions. As we show in Fig.~\ref{fig:qsd_meso_3}, increasing the value of $N$ provides better agreement with the Markov chain for all methods.

The most testing regime in which to challenge our various prescriptions for the noise is when the noise is strong and $\lambda$ is chosen such that there is a stable fixed point lying very close to one of the boundaries. We show results for this scenario in Fig.~\ref{fig:qsd_meso_4}. Here, the results from the random replacement method are easy to interpret. Returning a trajectory to a randomly chosen point in the unit interval has the effect of reducing the amount of time trajectories spend near the stable fixed point, causing a broadening of the probability distribution. In this figure it is clear that the censored distributions perform much better than the truncated distributions. This is because the renormalization of the truncated distribution means that if $z_t$ is close to zero then the following iteration, $z_{t+1}$, is artificially more likely to fall near (but not at) the boundary, since in this case the noise distribution is very large near zero, and renormalizing it adds even more weight to this region. Subsequent iterations, then, will very likely become `trapped' inside it. We also note that, for both truncated and censored distributions, the Gaussian distribution performs as well as the more complicated skew-normal distribution.

Although the method of censoring outperforms the other methods, it is clear from our results that some discrepancies are apparent when trajectories spend an appreciable amount of time very close to zero. We hypothesized that the key to the discrepancies apparent here, and in the other methods, is an important difference in the behavior of the microscopic and mesoscopic models. It is noticeable in Fig.~\ref{fig:qsd_meso_4} that the blue solid line, denoting the largest eigenvector of $\bm{\hat{Q}}$ (or, equivalently, the second-largest eigenvalue of $\bm{Q}$) does not extend all the way to $z=0$. This is because when we form $\bm{\hat{Q}}$ we remove the absorbing state present in $\bm{Q}$. So the first lattice point in the reduced microscopic system corresponds to $z=1/N$, not $z=0$. In contrast, in a mesoscopic model, a trajectory can enter the region $0 < z_t < 1/N$, and then later return to the region $1/N \leq z_t < 1$. This is then a possible explanation for why the region near zero is more `populated' than expected from the metastable state of a microscopic model with an absorbing boundary.

We tested this theory by using a modified censoring for the noise distribution. Previously, the probability mass lying outside the unit interval was placed at the boundaries. Now, we stipulate that the total probability outside the interval $1/N \leq z_t <1$ be placed at the boundaries. We used a Gaussian distribution for the noise, and will refer to this method as the \textit{modified censored Gaussian}. The bottom right panels in Figs.~\ref{fig:qsd_meso_1}--\ref{fig:qsd_meso_4} show the results obtained. A clear difference between the two types of censoring is visible in Fig.~\ref{fig:qsd_meso_4}. Although the modified censoring method does not perfectly capture the results from the microscopic system, the results are qualitatively more similar, in that the probability distribution has a local maximum. For the parameters chosen in Fig.~\ref{fig:qsd_meso_1} trajectories vary rarely visit the interval $0 < z_t \leq 1/N$, and so no difference between the types of censoring is visible. In Fig.~\ref{fig:qsd_meso_2}, the results from the modified censoring method are more similar to the behavior of the microscopic model. However, as $N$ increases, the difference between the methods diminishes, as the region $0 < z_t \leq 1/N$ becomes very small (see Fig.~\ref{fig:qsd_meso_3}).

\section{Discussion}
\label{sec:discuss}
In a series of earlier papers \cite{challenger_13,challenger_14,parra_14} we developed a formalism where, starting from a microscopic model, a mesoscopic description of a discrete-time stochastic process can be studied. Previously many authors had studied discrete-time stochastic models but they either assumed that time became continuous (as the number of states became large) or simply added a noise term to the deterministic dynamics. In contrast we began from a microscopic model with discrete states and discrete time, and derived the corresponding mesoscopic model now with a continuous state space and discrete time, which has a multiplicative noise generated by the underlying microscopic process. We found an analog of the Fokker-Planck equation, but which contained an infinite number of derivatives in the state variable. As discussed in Sec.~\ref{sec:meso}, this can be thought of as a reflection of the fact that large jumps can occur in systems where time is discrete.

It is possible to manipulate this partial differential equation into one with no more than second order derivatives and from this obtain an SDE. The advantage of the SDE is that it may be simulated for arbitrarily large $N$, which is not the case for the microscopic model. However, one difficulty arises: unlike in the continuous-time situation, the large jumps can lead to transitions which leave the state space on which the microscopic model is defined. It is clear that we have to somehow suppress these escape events, since the probability of the system being outside of the state space must be strictly zero, but the question is: how should this be imposed on the SDE?

In this paper we have investigated a number of aspects of this feature, leading up to a variety of algorithms designed to forbid escape trajectories, described in Sec.~\ref{sec:trun_cen}. After studying the results from five prescriptions for preventing trajectories leaving the interval, we propose that in the mesoscopic approximation an extra ingredient is needed which is not required when the time is continuous. This is that the noise term in the SDE should not have a distribution which is Gaussian but instead should be a censored Gaussian. The censoring is carried out in such a way that it ensures that any realization of the noise does not lead to a trajectory which leaves the interval on which the system is defined. This may appear to detract from the perceived simplicity of the SDE formalism, but we can make little analytical progress from the SDE, and so most investigations are numerical, for which the simplicity of the noise distribution is not so important. Furthermore, if we are only interested in the quasistationary distribution of the Markov chain, the censoring can be easily implemented by using a standard Gaussian distribution and discarding the trajectories that leave the interval in which the process is defined.

We compared the results found by using a censored Gaussian with results using other methods, namely (i) using a truncated distribution, (ii) replacing trajectories back in the interval in some random position once they leave, and (iii)--(iv) using a truncated and censored skew-normal distribution, which better approximates the continuous analog of the microscopic noise distribution. We also compared these results to the desired behavior obtained from the Markov chain for values of $N$ where this is feasible. When carrying out this comparison for the quasistationary probability distribution, we found only small variation between methods for parameter values when the map either stays far from the boundaries or explores almost the entire unit interval---the latter case requiring larger values of $N$ to obtain good agreement with the Markov chain---while in the case when the map has a stable fixed point close to one of the boundaries, only the censored distribution yields reasonably accurate results. These can be improved if the censoring is performed in such a way that the SDE is forbidden from entering the region $0<z<1/N$, which does not feature in the quasistationary distribution of the Markov chain.

It is important to note that the parameter choices explored in this paper, when comparing different methods of dealing with escape events, represent extreme cases of the behavior of the system. In practice, the SDE approach will be employed for larger values of $N$, for which it is impractical to analyze the Markov chain directly. In such cases, we would expect the mesoscopic quasistationary distribution to accurately reproduce its microscopic counterpart when employing a censored distribution for the noise. We would also expect the original censored distribution to become sufficiently accurate as $N$ increases, eliminating the need for the modified version. Since we find no noticeable difference between the Gaussian and skew-normal versions of the censored distribution, the simpler Gaussian implementation is advised.

The questions we have addressed in this paper will occur whenever simulations of SDEs such as Eq.~(\ref{stoch_diff_eqn}) are encountered. Although they have been investigated in the context of a model which has the logistic map as its deterministic limit, the prescription that we have discussed is applicable to more general systems, including systems in more than one dimension. We therefore hope that the results we have found will prove useful in the study of all mesoscopic discrete-time models.

\begin{acknowledgments}
CP-R was funded by CONICYT Becas Chile scholarship No.~72140425 and D.F. acknowledges financial support from H2020-MSCA-ITN-2015 project COSMOS 642563.
\end{acknowledgments}

\appendix

\section{Inclusion of higher-order jump moments}
\label{App_1}
In this appendix we provide details for constructing the characteristic function of the probability distribution for the noise, as displayed graphically in Sec.~\ref{sec:meso}. Our starting point is the definition of the jump moments, found in Eq.~(53) in Ref.~\cite{challenger_14}:
\begin{equation}
J_r(p)=\frac{1}{N^r}\sum_{s=0}^r {r\choose s}(-1)^{r-s}(Np)^{r-s}\mu_s(p),
\end{equation}
where we include a factor $(-1)^r$ which was missing in Ref.~\cite{challenger_14}. Here $\mu_s(p)$ is the $s^{\rm th}$ moment of the microscopic distribution, and is defined through the binomial structure of the transition matrix $\bm Q$, i.e., $\mu_s(p)=\sum_n n^sQ_{nm}$. From Eq.~(\ref{J_r}) we also have the relation
\begin{equation}
J_r(p)=\langle [z_{t+1}-p_t]^r \rangle_{z_t=z}=\langle \eta_t^r \rangle .
\end{equation}
We wish to use knowledge of the jump moments to reconstruct the probability distribution for the noise $\eta$. This distribution will in general depend on the value of $p_t$, so we label it $\Pi(\eta|p)$. This is carried out by using the characteristic function, defined as the Fourier transform of $\Pi(\eta|p)$:
\begin{equation}
\phi(s|p) \equiv \int_{-\infty}^{\infty} \textrm{d}\eta \Pi(\eta|p)\textrm{e}^{i\eta s}.
\end{equation}
Expanding the exponential we find
\begin{equation}
\phi(s|p) \equiv \int_{-\infty}^{\infty} \textrm{d}\eta \sum_{r=0}^\infty \frac{(is\eta)^r}{r!} \Pi(\eta|p)= \sum_{r=0}^\infty \frac{(is)^r}{r!}\langle \eta^r(p) \rangle .
\end{equation}
We can calculate the left-hand side of the above equation using our knowledge of the jump moments. For the figures presented in this paper, we used the first 150 jump  moments to approximate $\phi(s|p)$. We can then find the distribution $\Pi(\eta|p)$ by taking the inverse Fourier transform.

\section{Microscopic noise distribution}
\label{App_2}
In this second appendix we explore the properties of the microscopic distribution of which the noise distribution $\Pi(\eta|p)$, discussed in Appendix \ref{App_1}, is the mesoscopic equivalent.

We therefore return to the model defined as a Markov chain, and recall from the definition of the transition matrix $\bm Q$, that its $m$th column $Q_{nm}, n=0,1,\ldots,N$ represents the (discrete) probability distribution of jumps starting from the $m$th state. So this is in some sense the microscopic analog of $\Pi$. The main difference is that in the former the initial state $m$ is fixed, whereas in the latter it is $p=p(z)$ that is fixed. In addition, in the microscopic version, the initial state $m$ and the final state $n$ are, respectively, $Nz_t$ and $Nz_{t+1}$ in the notation of the mesoscopic formulation. However, $\Pi$ is more naturally expressed as a function of $p$ and $\eta$, and so the correct analog of $Q_{nm}$ is $\Pi(z_{t+1}|p)=\Pi(p + \eta|p)$. That is, it is the probability distribution of states $z_{t+1}$, given that $z_t$ satisfies $p(z_t)=p$, but expressed in terms of the variables $\eta$ and $p$.

We can now ask if $Q_{nm}$ has any properties which we would wish to impose on $\Pi(p + \eta|p)$. One significant feature can be seen by examining the probabilities of jumping from the given state $m$ to the boundaries at $n=0$ ($z=0$) and $n=N$ ($z=1$). These are given, respectively, by $Q_{0m} = (1-p)^N$ and $Q_{Nm} = p^N$, where $p = \lambda(m/N)[(N-m)/N]$ for the logistic map. These elements are nonzero except for $p=0$ and $p=1$, i.e.~for $m=0$ and $m=N$, and yet they are by definition $0$ for $n$ negative or greater than $N$. This suggests that we should ask that $\Pi$ does not smoothly tend to zero at the boundaries, but instead has discontinuities at $p + \eta = 0$ and $p + \eta = 1$, that is, at noise strengths $-p$ and $1-p$.

\end{document}